\documentclass[11pt,a4paper]{article}
\usepackage[left=3 cm, right=2.5 cm, top=3.5 cm, bottom=3.5 cm]{geometry} 
\usepackage[pdfborder={0 0 0},plainpages=false,pdfpagelabels]{hyperref}
\usepackage[latin1]{inputenc}
\usepackage{amsmath}
\usepackage{amsfonts}
\usepackage{amssymb}
\usepackage{makeidx}
\usepackage{graphicx}
\usepackage{float}  
\usepackage{multicol} 
\usepackage{subfig}
\begin{document}
\begin{center}
\textbf{MEASUREMENTS OF DOSE DISTRIBUTION OUTSIDE THE TREATMENT AREA IN CASE OF RADIOTHERAPY TREATMENT USING POLYSTYRENE PHANTOM}
\linebreak\linebreak  
Md. Farid Ahmed$^{1*}$, S. Roy$^{1}$, G. U. Ahmed$^{2}$ and F. K. Miah$^{1}$ \linebreak\linebreak
$^{1}$Health Physics Division, Atomic Energy Centre, Bangladesh Atomic Energy Commission, P. O. 
Box-164, Ramna, Dhaka-1000, Bangladesh. \linebreak\linebreak  
$^{2}$ Department of Physics, Bangladesh University of Engineering and Technology, Ramna, Dhaka-1000, 
Bangladesh.\linebreak\linebreak  
$^{*}$Present Address of Corresponding Author: \\ York University, 4700 Keele Street, Toronto, Ontario, Canada-M3J 1P3. \\ E-mail: mdfarid@yorku.ca
\end{center}
\bigskip
\textbf{ABSTRACT:}
Dose distribution (depthwise and laterally) to organs outside the radiotherapy treatment field can be significant and therefore is of clinical interest from the radiation protection point of view. In the present work, measurements were performed in a locally fabricated polystyrene phantom using TLD chips (LiF-100) for different teletherapy units ($^{60}$Co gamma ray, 120 kVp X-ray and 250 kVp X-ray) to estimate the dose distribution at distances up to 40 cm from the field edge along the central axes of the field size. Finally, the dose distribution for $^{60}$Co beam energy is parameterized  as a function of depth, distance from field edge, and field size and shape. 
\bigskip\\
\textbf{Shortened version of the title:}\\
A DOSIMETRY ASSESSMENT IN CASE OF RADIOTHERAPY TREATMENT 
\bigskip\\
\section{INTRODUCTION} In radiotherapy treatment, however, comparatively large exposure dose from X-rays or gamma rays is used in the treatment of cancer patients which not only kills the cancerous or malignant tissues but also, unfortunately, causes radiation exposure to surrounding healthy tissue and other critical organs like central nervous system, haemopoietic system, eye lens, gonad etc. The chance of a second cancer occurring following radiotherapy for a variety of cancers among long-term survivors is also well-known. In the United Nations report on the effects of atomic radiation (UNSCEAR, 1977) it is estimated that exposure of one million persons to one rad each of ionizing radiation will induce about 20 leukaemias and 100 fatal cancers of other sites. From the radiation protection point of view, one of the basic principle of using ionizing radiation in medical fields (Benson, 1990) is that the dose to surrounding tissues should be minimized by using the best available treatment planning and by taking measures to reduce the dose as far as possible to other parts of the body. Treatment planning should be extended to regions outside the immediate vicinity of the target volume, thus enabling calculation of radiation doses to other organs and tissues, which are adjacent to the treatment area for the purpose of estimating risk of complications. \\ \\ 
The physical part of treatment planning for external beam radiation therapy involves use of information on dose distribution. Information for treatment planning, including data on depth doses and dose distributions, as supplied by the equipment manufacturer, should not be used clinically without independent confirmation of the actual values (ICRP publication 44, 1985). It is thus, necessary to measure dose to different parts of the human body due to various radiotherapy procedures performed over cancer patients. Several studies were performed to measure the radiation dose outside the primary beam for square or rectangular fields, by applying different methods in many hospitals (Ahmed et al., 1999; Brandan et al., 1994; Francois et al., 1988; Green et al., 1983; Green et al., 1985; Kase et al., 1983; Ahmed, 1994; Miah et al., 1998; Sherazi et al., 1985; Starkeschall et al., 1983; Ahmed, M.F. 2000).  \\ \\
In a developing country like Bangladesh, we do not have sufficient data about the dose distribution over different organs of the body in different radiotherapy procedures which might give an idea about the existing situation in the country that might help take steps against hazardous effects (Ahmed et al., 1999). Therefore, the main objective of the present study is to determine the dose distribution over a phantom in the case of radiotherapy treatment of cancer patients. The overall measurements were conducted at square and rectangular fields, which are usually used in treatment of cancer patients throughout Bangladesh. Therefore, the measurements were performed in the context of Bangladesh to determine the depth factor (DF), and Elongation Factor (EF) over a polystyrene phantom in the $^{60}$Co teletherapy unit at the Delta Medical Center Limited, Dhaka, Bangladesh and Deep therapy X-ray unit at the Radiotherapy Department, Dhaka Medical College Hospital, Dhaka, Bangladesh. The Thermoluminescent Dosimeter (TLD) chips (TLD-100) were employed for the determination of dose. The measured data would be useful to recommend measures for the protection of other organs of cancer patients undergoing radiotherapy, which is one of the major objective in radiological practice.  
\section{INSTRUMENTATION, MATERIAL AND METHOD}
\subsection{TLD System}
In the present study, Lithium Fluoride (LiF) with impurity doping in the form of chips having commercial names of TLD-100 (natural isotopes with ratio of 7.5\% $^{6}$Li and 92.5\% $^{7}$Li and of size 1/8 inch $\times$ 1/8 inch $\times$ 0.035 inch and weighing about 24 mg) have been used as TL dosimeters and the chips were supplied by the Harshaw Chemical Company, Cleveland, Ohio, USA. The TLD chips were read with the help of Harshaw TLD system, Model 3500 manual TLD reader. This TLD system is shown in Fig. 1(a).
\begin{figure}[h]
\includegraphics[width=1\textwidth]{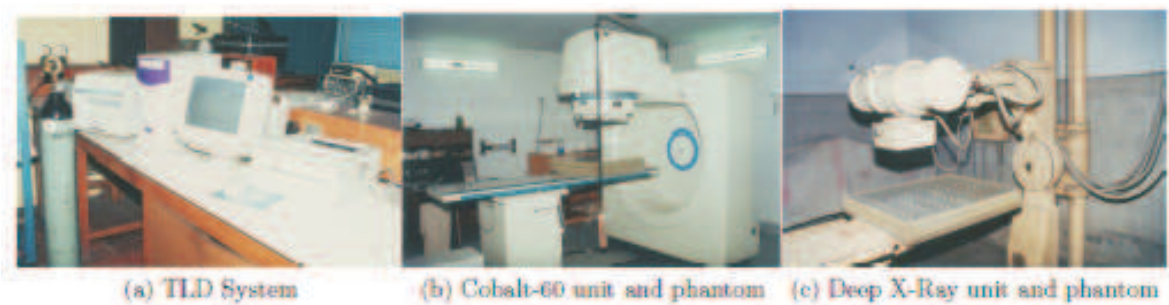} 
\caption{Pictures of different instruments were used for the measurements.}
\end{figure}
The grouping of the chips under study were done using irradiator of a $^{90}$Sr/$^{90}$Y source (Model 2210, Nominal Activity: 33 MBq (0.917 mCi, 18 November 1998), which was calibrated by Bicron Technologies. The averaged $^{137}$Cs equivalent dose was found to be 6.90 mSv for 100 revolutions ($^{90}$Sr/$^{90}$Y irradiator calibration certificate). The Reader Calibration Factor (RCF) is the factor which converts TL signal (i.e., current in nC) into dose and the RCF of the TLD reader was found to be 0.007 mSv/nC and the Elemental Correction Coefficient (ECC) is the property of a chip, which is used to get the accurate radiation dose. Therefore, the individual TLD chips were determined using the $^{90}$Sr$/^{90}$Y irradiator. The RCF and ECC of the TLD reader and chips respectively were determined using the measured $^{90}$Sr/$^{90}$Y irradiator. 
\subsection{Phantom}
It is seldom possible to measure dose distribution directly in patients treated with radiation. Data on dose distribution are almost entirely derived from measurements in phantoms, tissue equivalent materials, usually large enough in volume to provide full-scatter conditions for the given beam (Fiaz, 1994). These basic data are used in a dose calculation system devised to predict dose distribution in an actual patient. \\ \\
In the present work, polystyrene slabs were used as a phantom material as shown in Fig. 1(c) where the phantom was on the couch of the X-ray unit. Precisely drilled holes on the polystyrene sheet along the X-axis/Y-axis diverging from the end of the field area with the dimension of 70 cm $\times$ 60 cm $\times$ 13 cm was used as the phantom. The variation of atomic composition of commercially available polystyrene with the human body has been shown to be small (Kase et al., 1983; Schulz et al., 1979). Therefore this material was employed in this study as a phantom without a specific chemical analysis of their atomic composition. 
\subsection{Method of Calibration for TLD chips}
For our present work, we used a $^{60}$Co teletherapy (ALCYON II, CGR, MeV, France) unit of activity 223.6 TBq (on 09 June, 1994) as shown in Fig. 1(b) and a Deep therapy X-ray unit (Siemens Ltd., Germany) as shown in Fig. 1(c). \\ \\ 
\begin{figure}[h]
\centering
\subfloat[For $^{60}$Co: $y=0.55x$]{\includegraphics[height=2 in, angle=270]{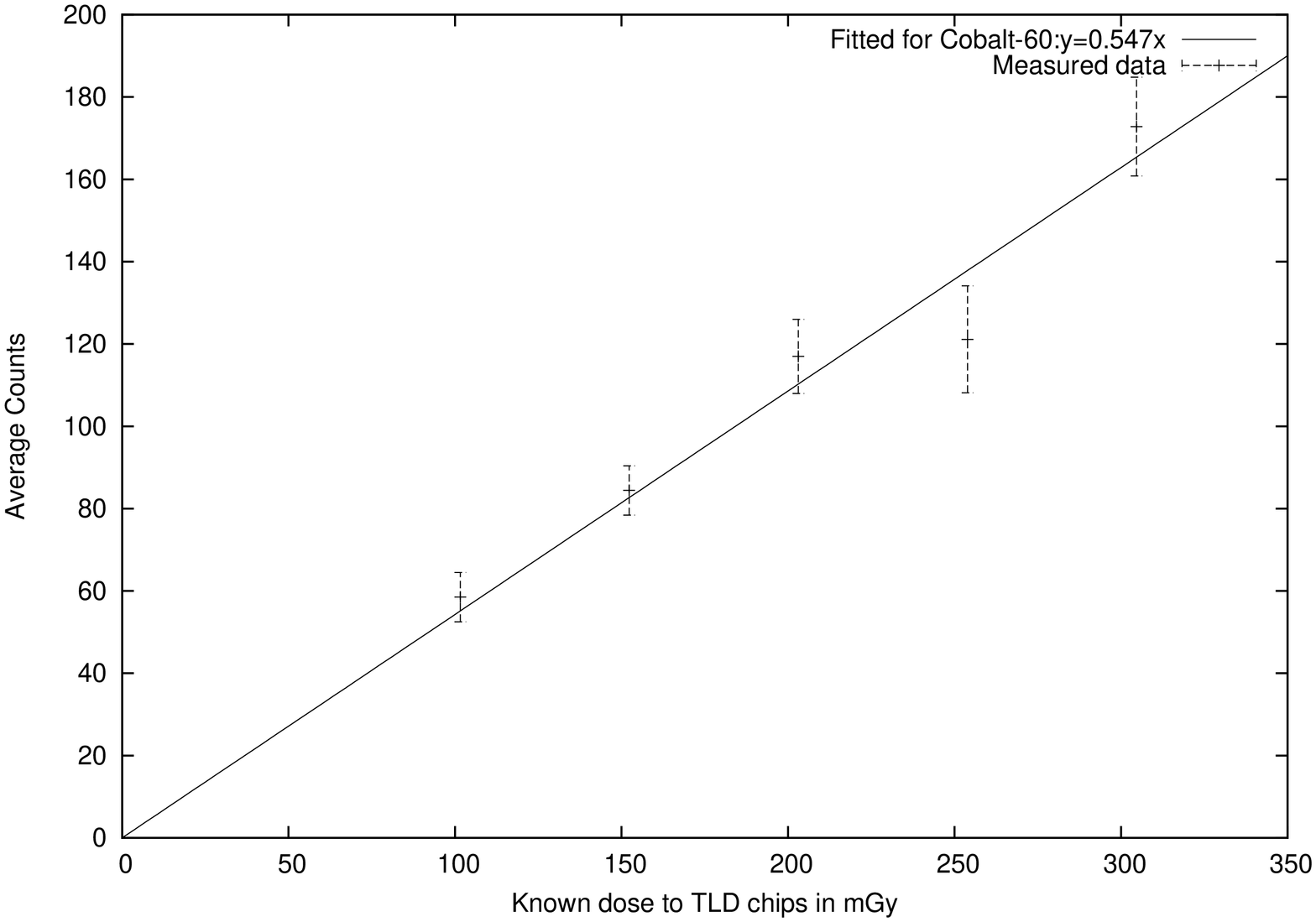}}
\subfloat[For 250-kVp X-ray: $y=0.67x$ ]{\includegraphics[height=2 in, angle=270]{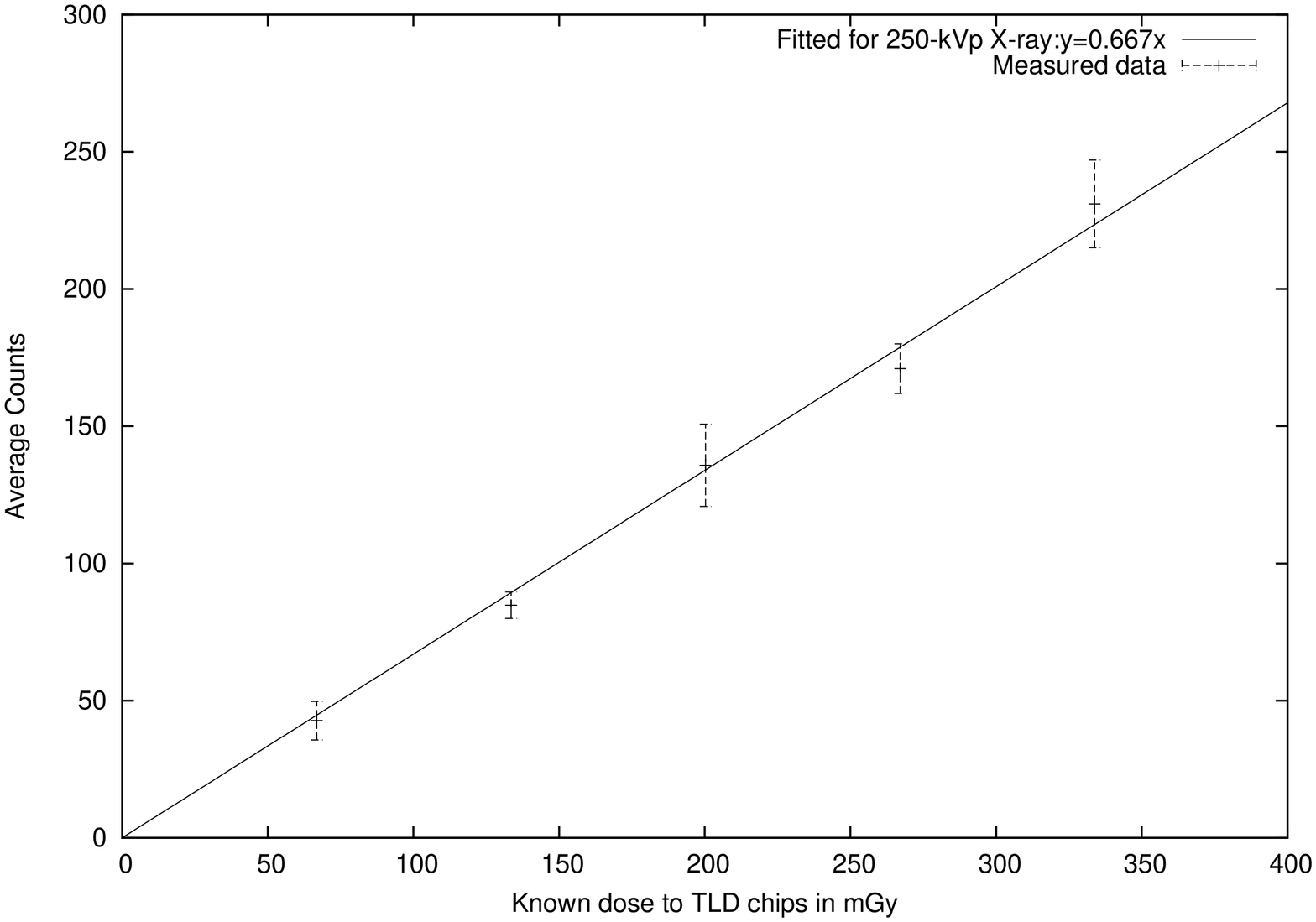}} 
\subfloat[For 120-kVp X-ray: $y=0.53x$]{\includegraphics[height=2 in, angle=270]{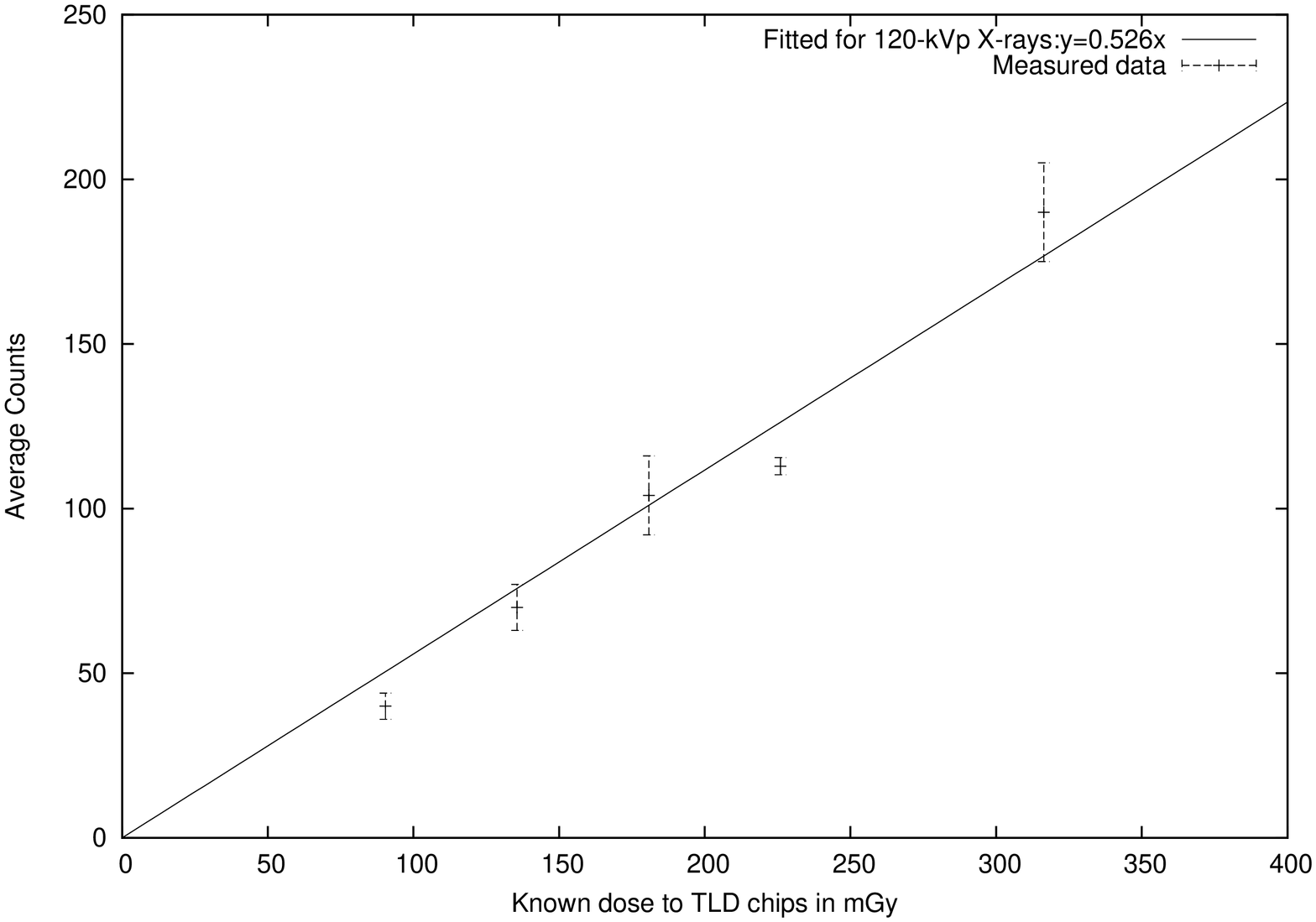}}
\caption{Calibration graphs (as well as equations) of average counts in a TLD reader to known dose to TLD chips (in mGy).}
\label{fig:EcUND4}
\end{figure}
In order to calibrate the TLD chips for a $^{60}$Co teletherapy unit, twenty-five (25) TLD chips were divided into five groups having five chips in each group. These five groups were irradiated by the $^{60}$Co teletherapy unit at a known dose of 101.55 mGy, 152.25 mGy, 203.1 mGy, 253.9 mGy and 304.65 mGy. The dosimetry of the $^{60}$Co unit was performed by using an ion chamber (PTW, UNIDOS, SI. No. N23323-3713, Vol. 0.3 cc) and electrometer (PTW, UNIDOS, Germany, SI. No. 10005 - 50146) procured from PTB, Braunsech, Weig., Germany. These chips were read in the TLD reader applying the corresponding Element Correction Coefficients (ECCs) and Reader Calibration Factors (RCFs). The calibration curve after performing a least square fit of the plotted data following a procedure given in (Gottfried, 1979) for $^{60}$Co teletherapy is shown in Fig. 2(a).  \\ \\
For Orthovoltage and Superficialvoltage Teletherapy (250 kVp, 120 kVp), fifty (50) field dosimeters were divided into ten groups having five chips in each group. Five groups were irradiated by 250 kVp X-ray at a known dose of 66.8 mGy, 133.5 mGy, 200.3 mGy, 267.1 mGy, and 333.8 mGy. Another five groups were irradiated by 120 kVp X-ray at a known dose of 90.4 mGy, 135.6 mGy, 180.8 mGy, 226.0 mGy, 316.6 mGy. The dosimetry of the orthovoltage and superficial voltage X-ray teletherapy were carried out by using a 0.6 cc capintec ion chamber and Farmer electrometer. These chips were read in the TLD reader applying the ECCs and RCF. The calibration curves after performing a least square fit of the plotted data following a procedure given in (Gottfried, 1979) for orthovoltage (250 kVp) and superficialvoltage (120 kVp) teletherapy are shown in Fig. 2(b) and 2(c) respectively. \\ \\ 
More detailed data on the procedure of the calibration of the TLDs, as reported elesewhere ( Ahmed et al., 1999; Goil De Planque Burke, 1976; Hussain et al., 1982; Md. Farid Ahmed, 1994; Md. Farid Ahmed 2000; User's Manual, 1993).
\subsection{Calculation Method}
Francois et al. (1988) derived a formula giving the dose on the X-axis/Y-axis of a rectangular field with an 
elongation coefficient $ec_{x}$/$ec_{y}$ on that axis, at a distance $d_{x}$/$d_{y}$ from the edge and 
at depth z in the tissue. The general expression of that dose is given by the following equation:
\begin{equation}
Dr_{i}(ec_{i},z,d_{i})=Ds(z=5, d_{i}) EF_{i}(ec_{i}, d_{i}) DF(z, d_{i}), i=x, y
\end{equation}
where, $Dr_{i}(ec_{i}, z, d_{i})$ is the dose at distance $d_{i}$ from the edge of the rectangular field (on the axis) with an elongation coefficient $ec_{i}$ and at depth $z$. $Ds(z=5, d_{i})$ is the dose of the square field $(ec=1)$ of equal area at the same distance (on the axis) at 5 cm depth. $EF_{i}(ec_{i},d_{i})$ is an elongation factor depending upon the elongation coefficient $ec_{i}$ of the rectangular field at the distance $d_{i}$ from the field edge on the $i(x,y)$-axis. $DF(z,d_{i})$ is a depth factor depending upon the depth of calculation $z$ at the distance $d_{i}$ from the field edge.
\subsubsection{Elongation Factors (EF)}
The Elongation factor $EF(ec_{i},d_{i})$ is defined (Francois et al., 1988) as followis:\begin{equation}
EF_{i}(ec_{i},d_{i})=\frac{Dr_{i}(ec_{i},z=5,d_{i})}{Ds(z=5,d_{i})}, i=x,y,
\end{equation}
where, $Dr_{i}(ec_{i},z=5,d_{i})$ is the dose at distance $d$ from the edge of a rectangular field with an elongation coefficient $ec$, at 5-cm depth. $Ds(z=5,d_{i})$ is the dose at the same distance and at the same depth for a square field of the same area.
\subsubsection{Depth Factors (DF)} The Depth factor $DF(z,d_{i})$ is defined (Francois et al., 1988) as follows:
\begin{equation}
DF(z,d)=\frac{Ds(z,d)}{Ds(z=5,d)}
\end{equation}
where, $Ds(z,d_{i})$ is the dose at distance $d_{i}$ from the edge of a square field of surface at depth $z$. $Ds(z=5,d_{i})$ is the dose at the same distance from the same field at 5-cm depth.
\subsection{Procedure for the Determination of Dose}
In the present work, the previously selected seventy-five (75) TLD chips of LiF-100 known as Field Dosimeters were of nearly the same ECCs (upper ECC limit 1.20 and lower 0.80) and were used for the measurement of dose over the phantom throughout the experiment. The ECC and identity of each chip was kept same throughout the experiment. \\ \\ 
In order to measure the distribution of dose outside the radiation beam, field dosimeters were distributed within precisely drilled holes on the polystyrene sheet along the X-axis/Y-axis diverging from the end of the field area. In this study, the scattering doses were measured at distances of 5 cm, 10, cm, 20 cm, 25 cm, 30 cm, 35 cm and 40 cm away from the end of the treatment area along the axis of the field. The field sizes (15 cm $\times$ 15 cm); (10 cm $\times$ 10 cm) and (5 cm $\times$ 5 cm) were selected for $^{60}$Co and X-ray teletherapy units.\\ \\  
In the case of the elongation factor, dose distribution was measured with field sizes (5 cm $\times$ 20 cm); (7.07 cm $\times$ 14.14 cm); (10 cm $\times$ 10 cm); (14.14 cm $\times$ 7.07 cm) and (20 cm $\times$ 5 cm), for the $^{60}$Co unit in order to maintain the elongation coefficient (EC) 4, 2, 1, 0.5 and 0.25 respectively. The field sizes were defined at the surfaces of the phantom with the source to surface distance (SSD) 80 cm. The points of interest for dose measurement were chosen at different distances, such as, from 10 cm to 40 cm with 5 cm interval, from the edge of the surface of the phantom. \\ \\
The dose distribution was measured at depths of 1 cm, 5 cm, and 10 cm and field size 10 cm $\times$ 10 
cm for the determination of the depth factor. The field sizes were defined at the surface of the phantom. The 
locations of dose measurements were chosen as in the previous set up. \\ \\
The treatment area, i.e., desired field size, was irradiated by 500 cGy and the TLD chips, which were irradiated by scattered radiation, were then read by the TLD reader and the dose data was calculated using corresponding calibration curve as shown in Fig. 2. The chips were then annealed following standard procedure (400$^{\circ}$C for 1 hour and 100$^{\circ}$C for 2 hours) and became ready for reuse. The overall measurements were conducted with the fields and depths, which are usually adopted in treatment of cancer patients throughout Bangladesh. \\ \\
The net counts of the TLD chips attributed to the exposure of scattered radiation due to radiotherapy were obtained following the procedure, as reported elsewhere (Ahmed et al., 1999; Ahmed, M. F., 1994; Ahmed, M. F., 2000; User's Manual, 1993). These counts were converted to dose in mGy employing the calibration graph in Fig. 2 represented by the least square fitted equations. 
\section{RESULTS}
The variation of dose at different distances with different square fields outside the beam for $^{60}$Co, 250 kVp 
X-rays, 120 kVp X-rays have been presented in Figs. 3 - 5. The given dose at the square fields were 500 cGy 
for $^{60}$Co and 500 R for X-rays. 
\begin{figure}[h]
\centering
\subfloat[Field-Size 15$\times$15 cm$^{2}$]{\includegraphics[height=2 in, angle=270]{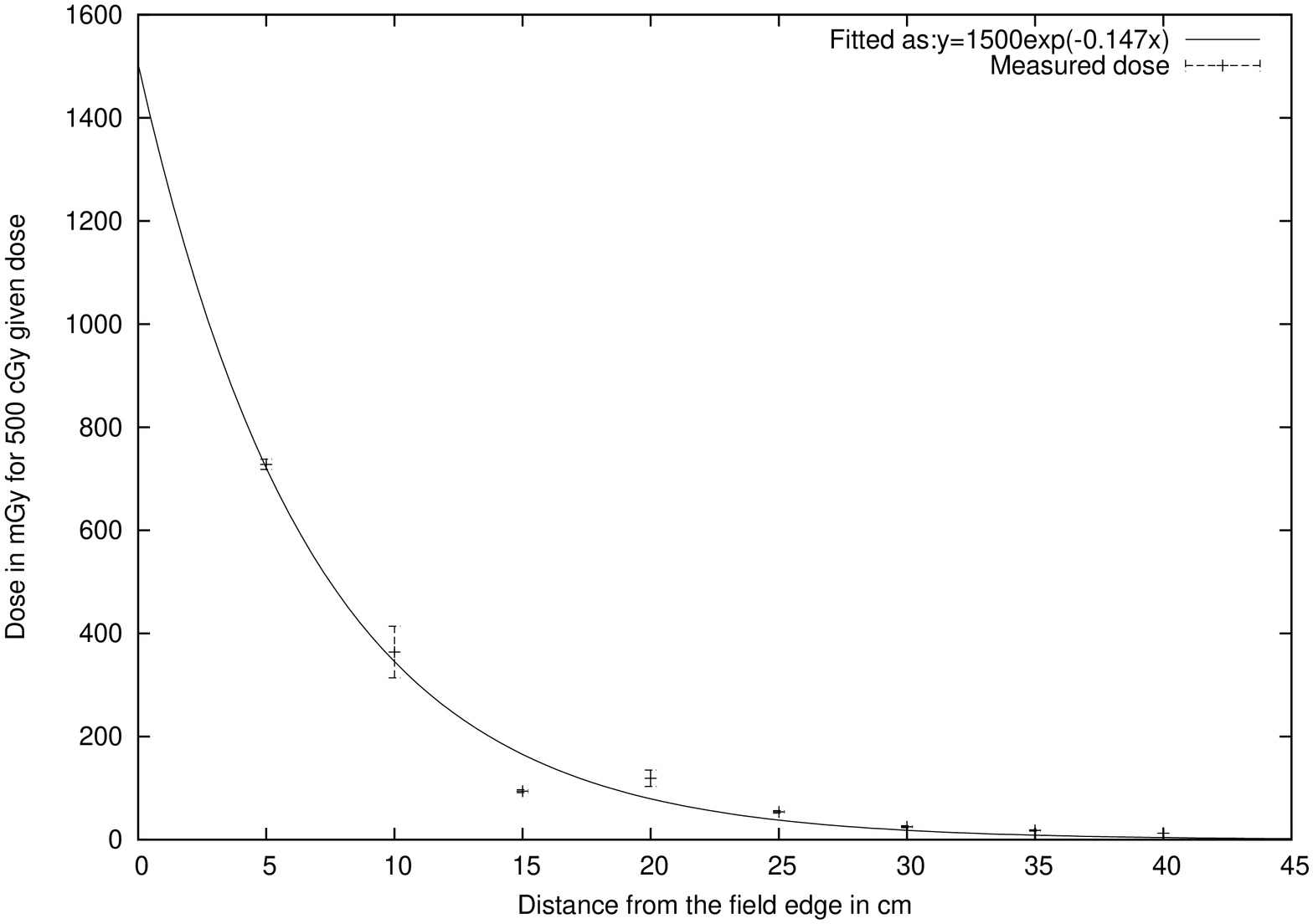}} 
\subfloat[Field-Size 10$\times$10 cm$^{2}$]{\includegraphics[height=2 in, angle=270]{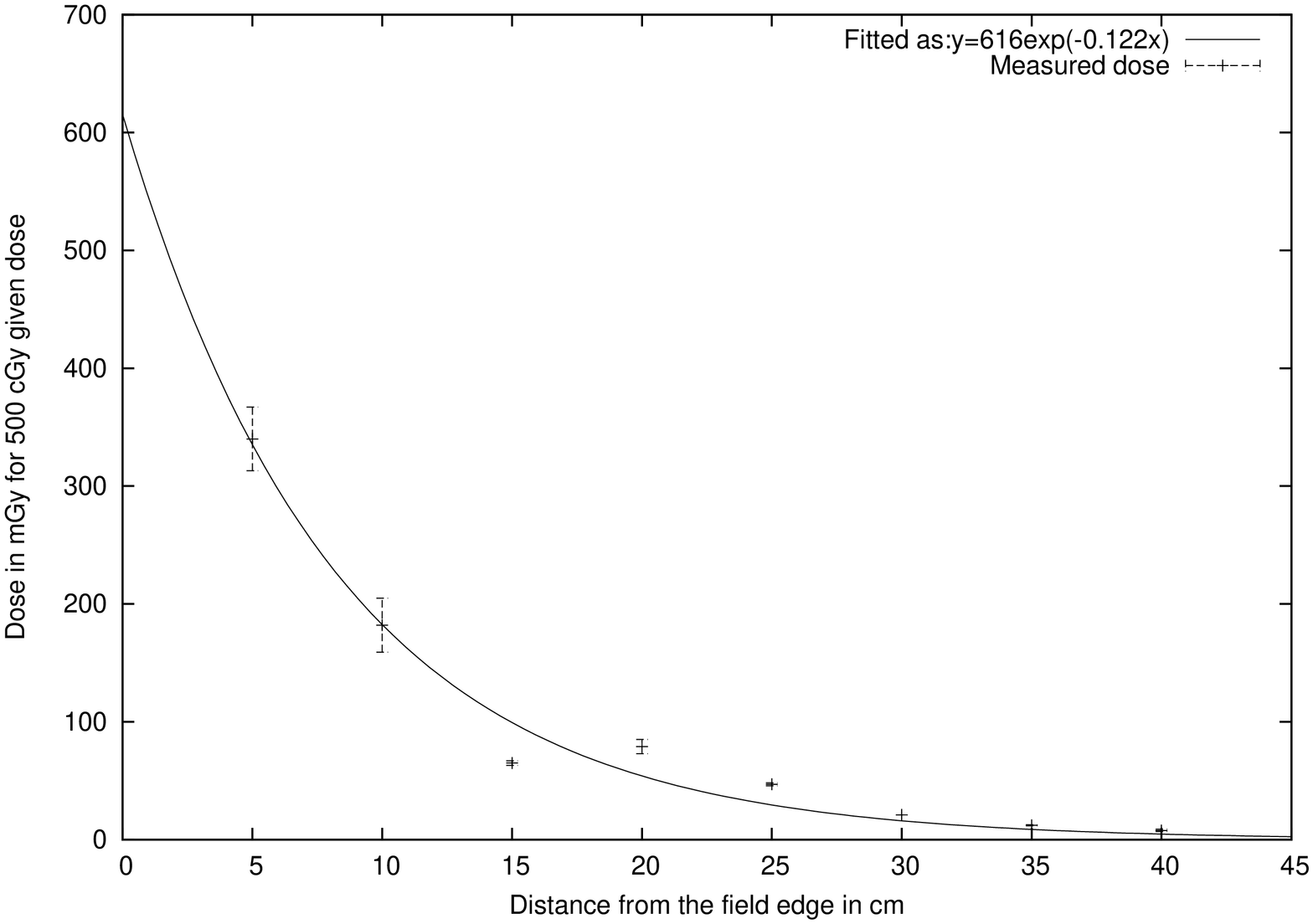}} 
\subfloat[Field-Size 5$\times$5 cm$^{2}$]{\includegraphics[height=2 in, angle=270]{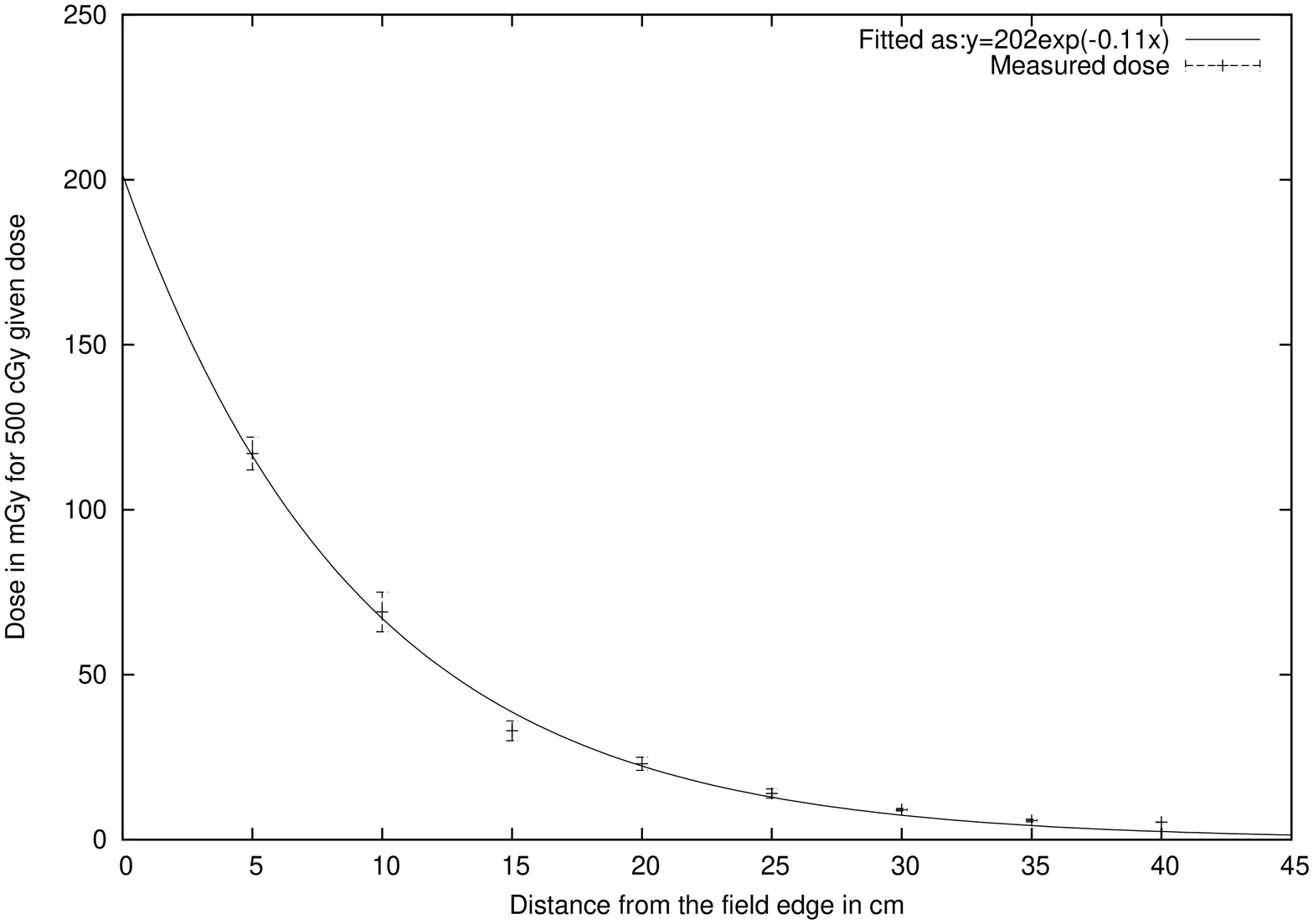}}
\caption{Dose distribution outside the beam for a $^{60}$Co unit with 80 cm SSD and 500 cGy given dose.}
\label{fig:EcUND2}
\end{figure}
\begin{figure}[h]
\centering
\subfloat[Field-Size 15$\times$15 cm$^{2}$]{\includegraphics[height=2 in, angle=270]{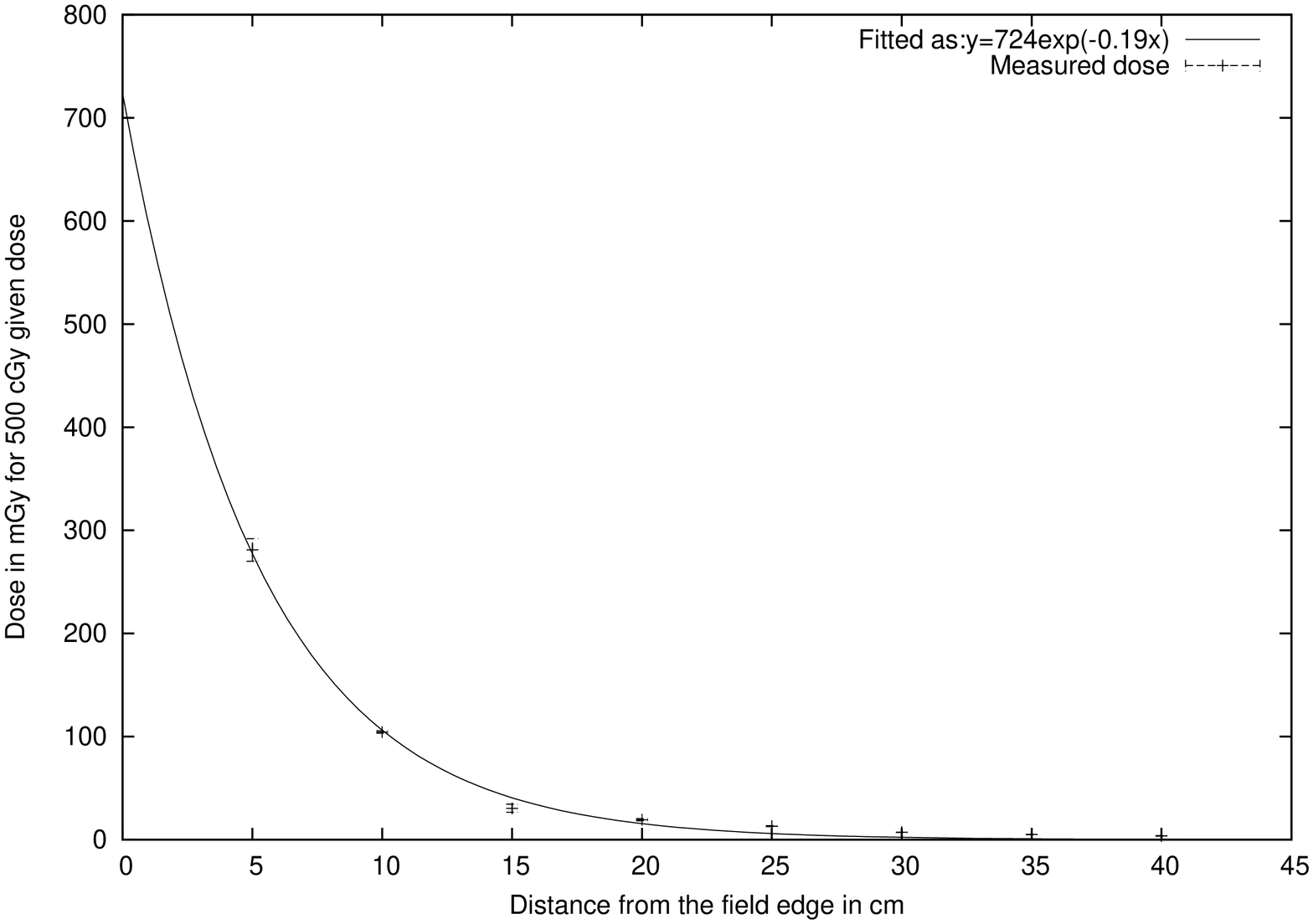}} 
\subfloat[Field-Size 10$\times$10 cm$^{2}$]{\includegraphics[height=2 in, angle=270]{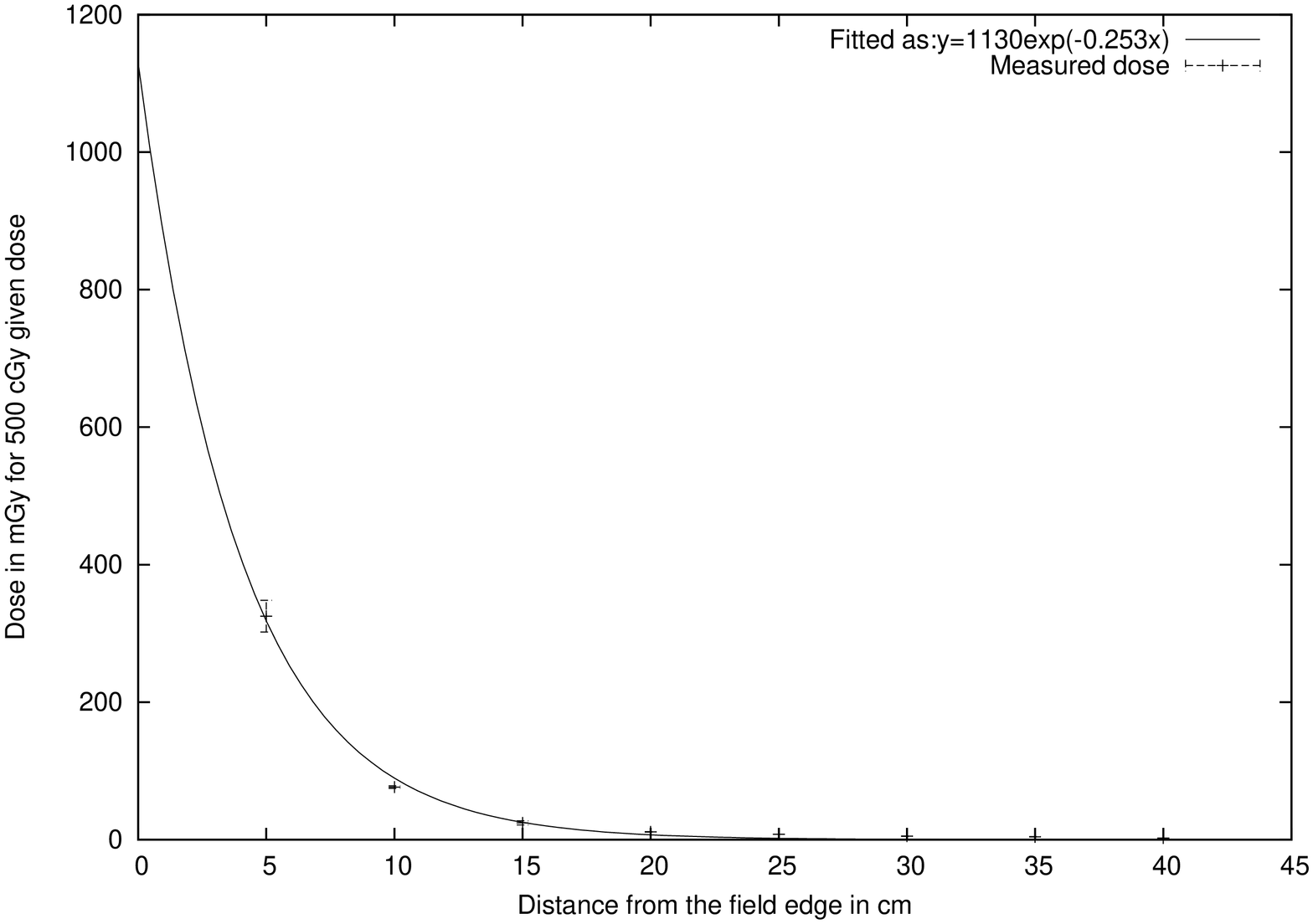}} 
\subfloat[Field-Size 5$\times$5 cm$^{2}$]{\includegraphics[height=2 in, angle=270]{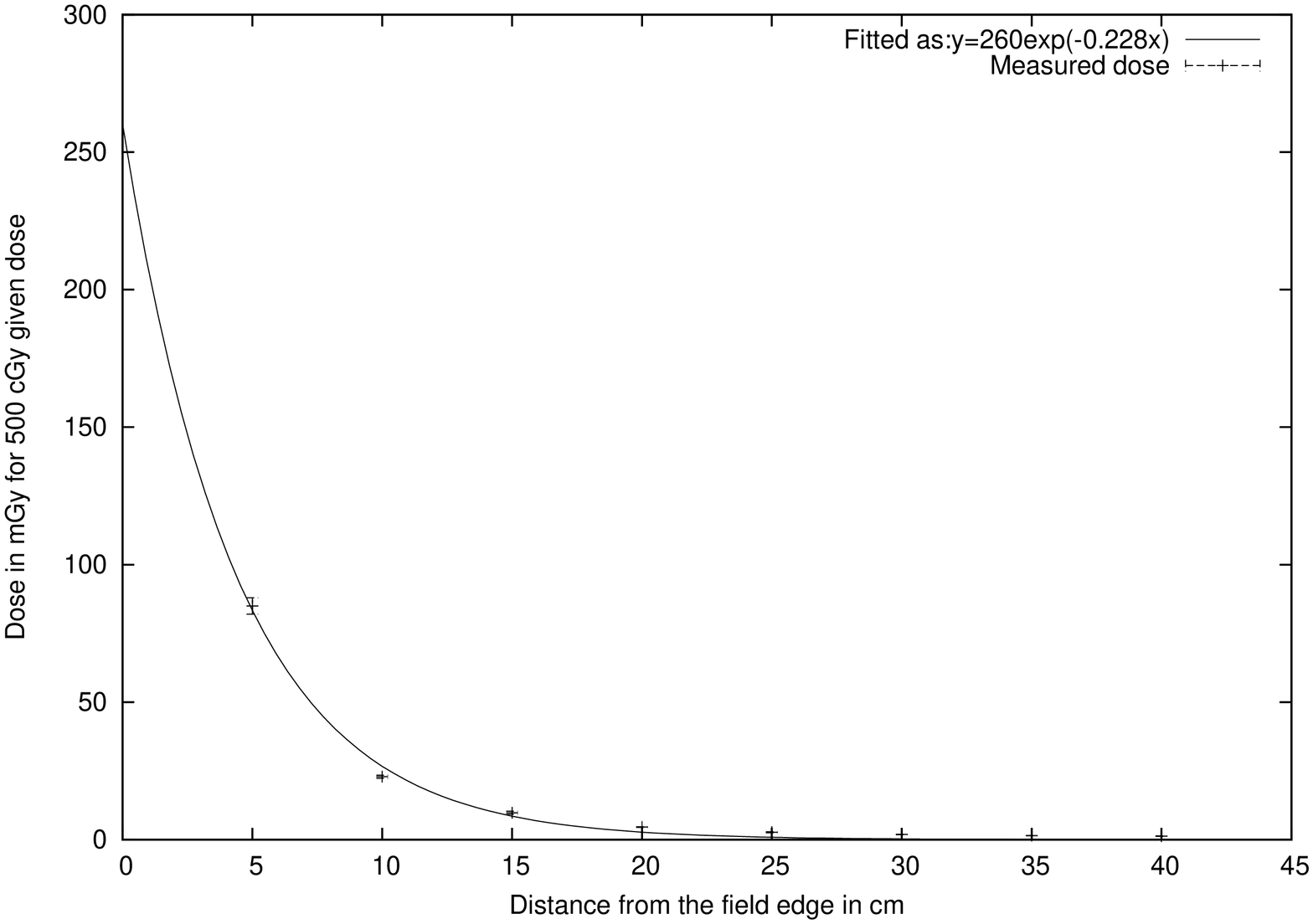}}
\caption{Dose distribution outside the beam for the 250 kVp X-ray unit with 50 cm SSD, 1 mm Cu filter and 500 R given dose.}
\label{fig:EcUND2}
\end{figure}
\begin{figure}[h]
\centering
\subfloat[Field-Size 15$\times$15 cm$^{2}$]{\includegraphics[height=2 in, angle=270]{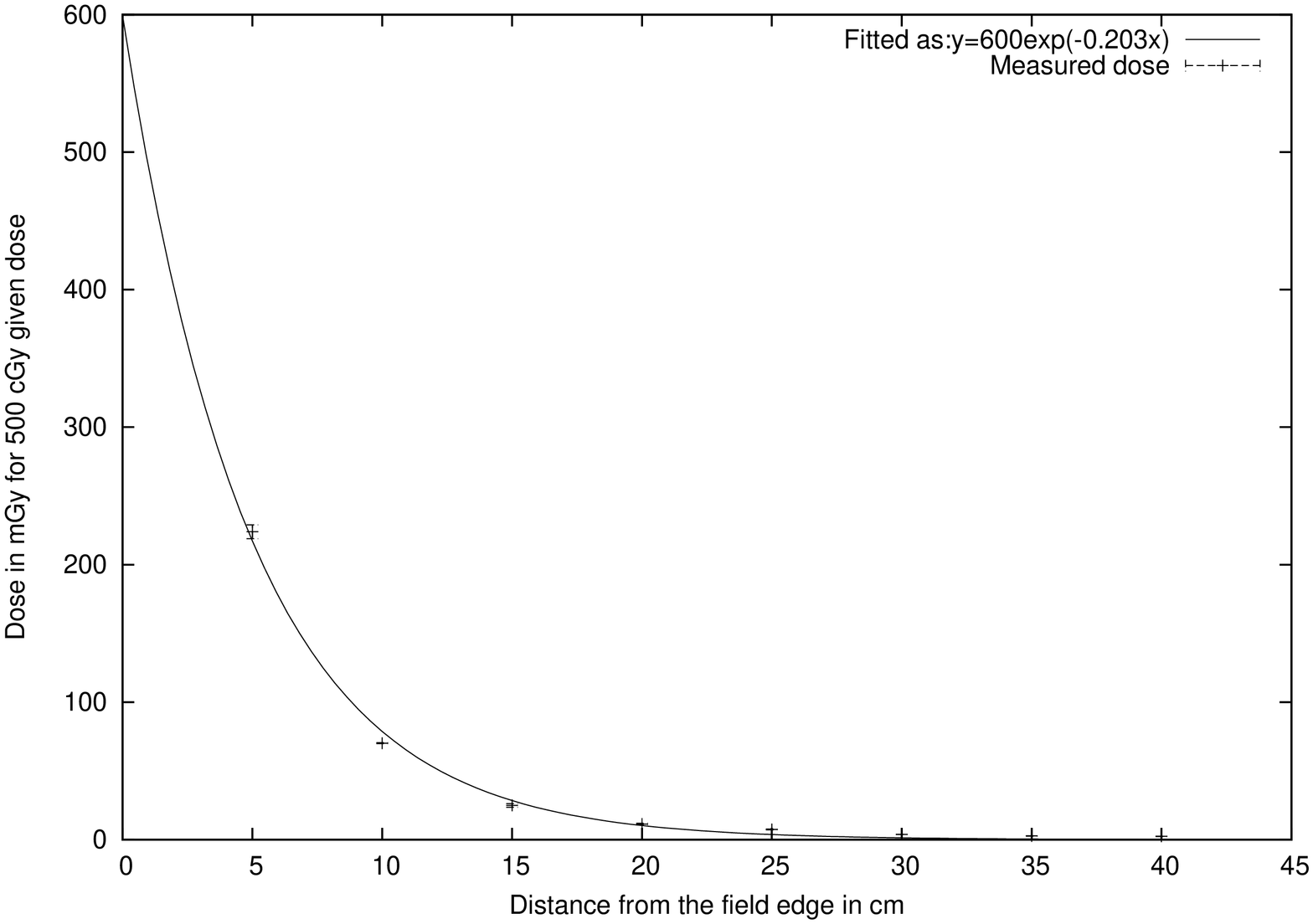}} 
\subfloat[Field-Size 10$\times$10 cm$^{2}$]{\includegraphics[height=2 in, angle=270]{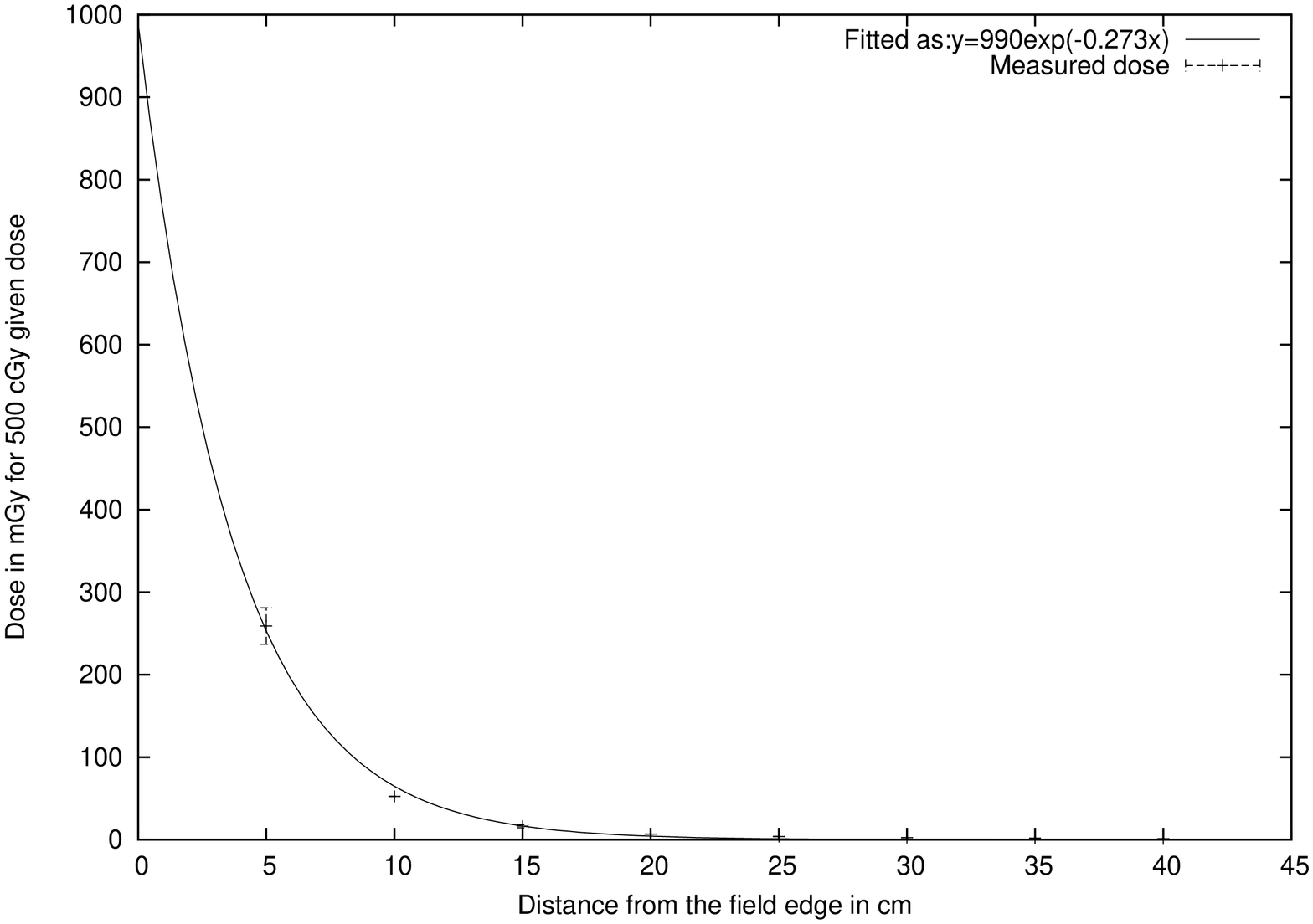}} 
\subfloat[Field-Size 5$\times$5 cm$^{2}$]{\includegraphics[height=2 in, angle=270]{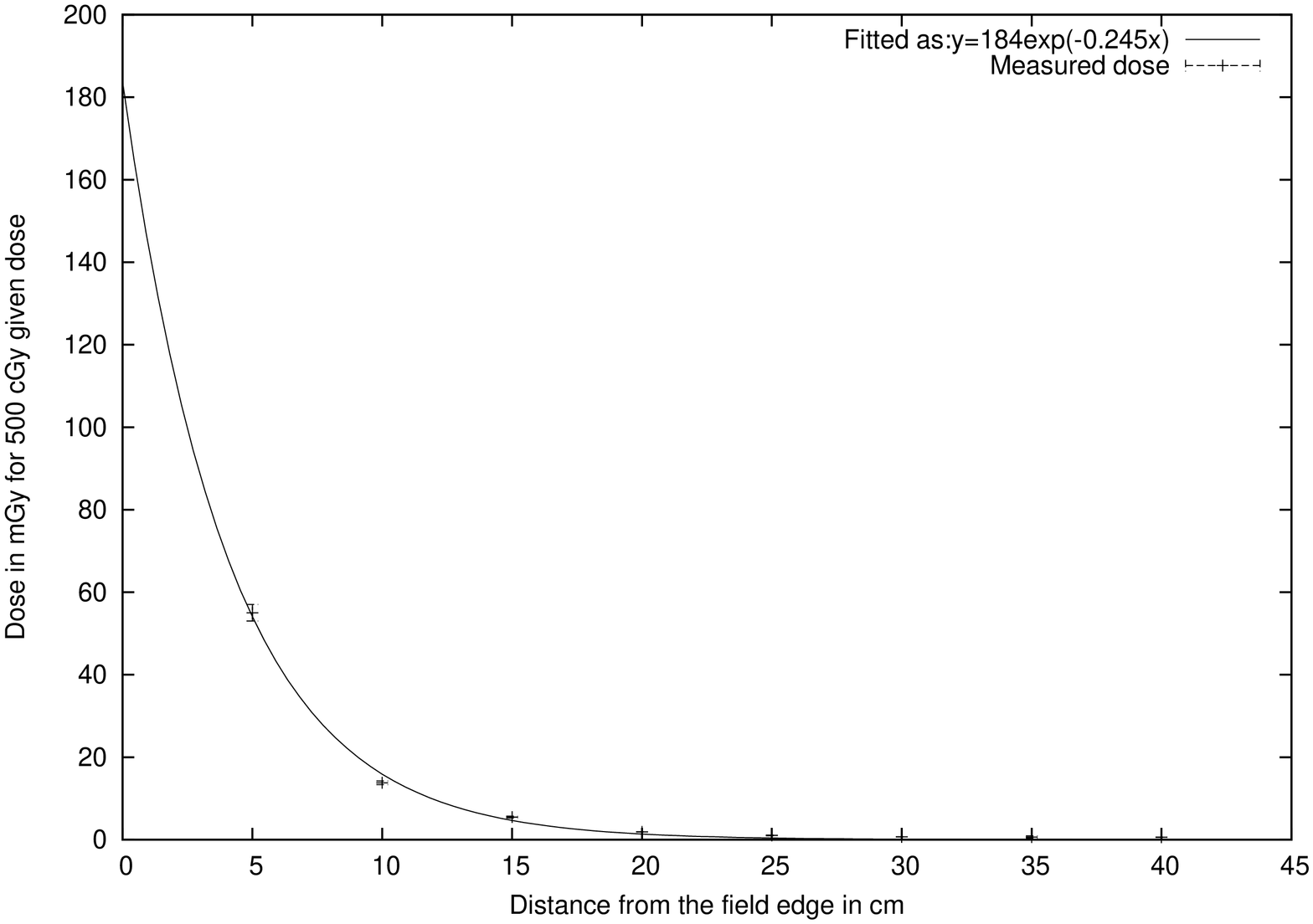}}
\caption{Dose distribution outside the beam for the 120 kVp X-ray unit with 35 cm SSD, 4 mm Al filter and 500 R given dose.}
\label{fig:EcUND2}
\end{figure}
Fig. 6 and Table 1 represent the measured dose distribution at different distance of the phantom due to $^{60}$Co for different field shapes (but same field area) i.e., for different elongation coefficients. It represents the variation of dose outside the beam with different elongation coefficients of a field area 100 cm$^{2}$. Based on our experience in fitting the data used in Figs. 3-5, the data of the Fig. 6 can be fitted closely to an exponential $"y=a \times exp(-bx)"$. Then using this fitted exponential equation, the data were recalculoated for corresponding elongation factors which are presented in Fig. 7. \\ \\
 \begin{center}
 \begin{tabbing}
Table 1: \= Measured dose (mSv) distribution at different distance of the phantom due\\ \> to radiotherapy 
 ($^{60}$Co teletherapy) for different field sizes (same field area). 
\end{tabbing} 
 [SSD: 80 cm; Given dose: 500 cGy; Depth: 5 cm] \linebreak
    \begin{tabular}{| l | l | l | l | l | p{3cm} |}
    \hline
    Distance & Dose for & Dose for & Dose for & Dose for & Dose for \\ 
    (d$_{i}$)  & 5 $\times$ 20 & 7.07 $\times$ 14.14 & 10 $\times$ 10 & 14.14 $\times$ 7.07 & 20 $\times$ 5 \\ 
    (cm) & (cm$^{2}$) & (cm$^{2}$) & (cm$^{2}$) & (cm$^{2}$) & (cm$^{2}$) \\  \hline
    10 & 58 $\pm$ 5 & 62 $\pm$ 2 & 64 $\pm$ 1 & 47 $\pm$ 9 & 40 $\pm$ 4  \\ \hline
    15	& 41 $\pm$ 3 &	33 $\pm$ 2 &	 31 $\pm$ 2 &	27 $\pm$ 6 &	23 $\pm$ 4	  \\ \hline
    20 &	27.5 $\pm$ 5.6 &	23 $\pm$ 1.6 & 19.8 $\pm$ 1.4	&13.3 $\pm$ 1.6 & 11.5 $\pm$1 \\ \hline
    25 &	19 $\pm$9&16.4 $\pm$0.8&	12.6 $\pm$2.1&	9.4 $\pm$0.7&8 $\pm$1.3 \\ \hline
    30	&	13.2 $\pm$	5.2&10.5 $\pm$1.5 &	7.9 $\pm$0.7&	6.1 $\pm$0.3&	4 $\pm$	0.2 \\ \hline
    35	&	8.3 $\pm$1.7 &	6.2 $\pm$ 0.6 &	5.3 $\pm$	0.2 & 3.7 $\pm$ 0.4 &	2.7 $\pm$ 0.04 \\ \hline
    40	&	5.2 $\pm$ 0.4 &	4 $\pm$ 0.7&3.3 $\pm$	0.3& 2.7 $\pm$ 0.5& 2.1 $\pm$ 0.09 \\ \hline
    \end{tabular}
\end{center}
\begin{center}
 \begin{tabbing}
Table 2: \= Measured dose (mGy) distribution at different distance of the phantom due\\ \> to radiotherapy 
 ($^{60}$Co teletherapy) for different depths with 10 $\times$ 10 cm$^{2}$ \\ \> field size. 
\end{tabbing} 
 [SSD: 80 cm; Given dose: 500 cGy] \linebreak
    \begin{tabular}{ | l | l | l | p{3cm} |}
    \hline
    Distance & Dose for & Dose for & Dose for  \\ 
    (d$_{i}$) (cm) & depth 1 cm & depth 5 cm & depth 10 cm \\  \hline
	10	&	55$\pm$ 0	& 59$\pm$ 1.6 &	45.95$\pm$ 1.05  \\ \hline   
    15 & 37$\pm$0.2 & 31.5 $\pm$2.2 & 28.9$\pm$ 2 \\ \hline 	
    20 & 29.3$\pm$0.9 & 22.2$\pm$ 0.5& 14.65$\pm$ 0.25 \\ \hline
    25 & 20.7$\pm$ 0& 15.5$\pm$ 0.2& 8.5$\pm$ 0.6 \\ \hline
    30 & 13.1$\pm$ 0.5 & 10.05$\pm$0.15 & 5.9$\pm$0.4 \\ \hline
    35 & 8.55$\pm$ 0.05 & 7.15$\pm$ 0.65 & 4.2 $\pm$ 0 \\ \hline
    40 & 4.3$\pm$0.2 & 3.3$\pm$0.1 & 3$\pm$ 0.1 \\ \hline
    \end{tabular}
\end{center}	
\begin{figure}[h]
\centering
\subfloat[Field Size 5 cm $\times$ 20 cm]{\includegraphics[height=3 in, angle=270]{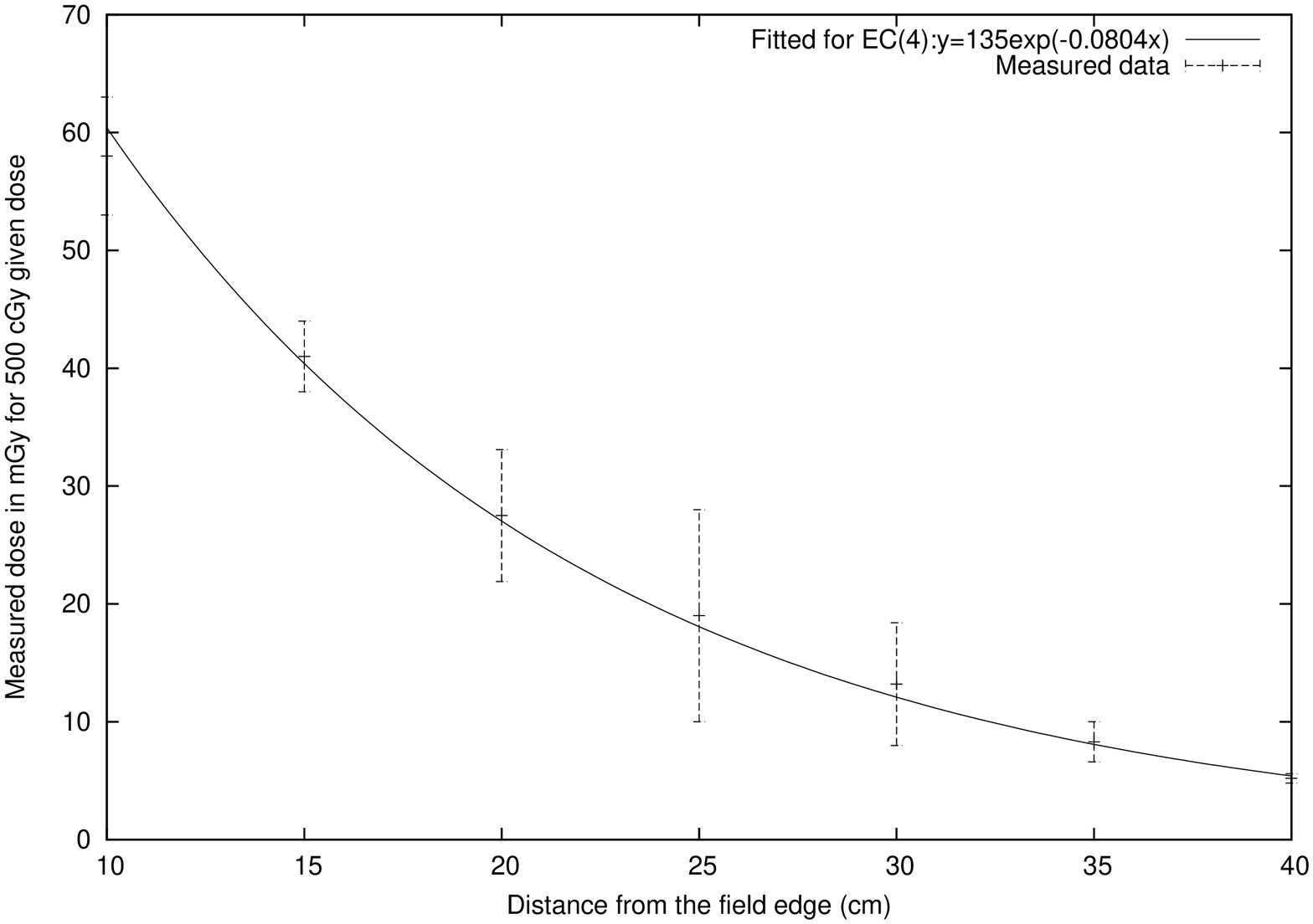}}
\subfloat[Field Size 7.07 cm $\times$ 14.14 cm]{\includegraphics[height=3 in, angle=270]{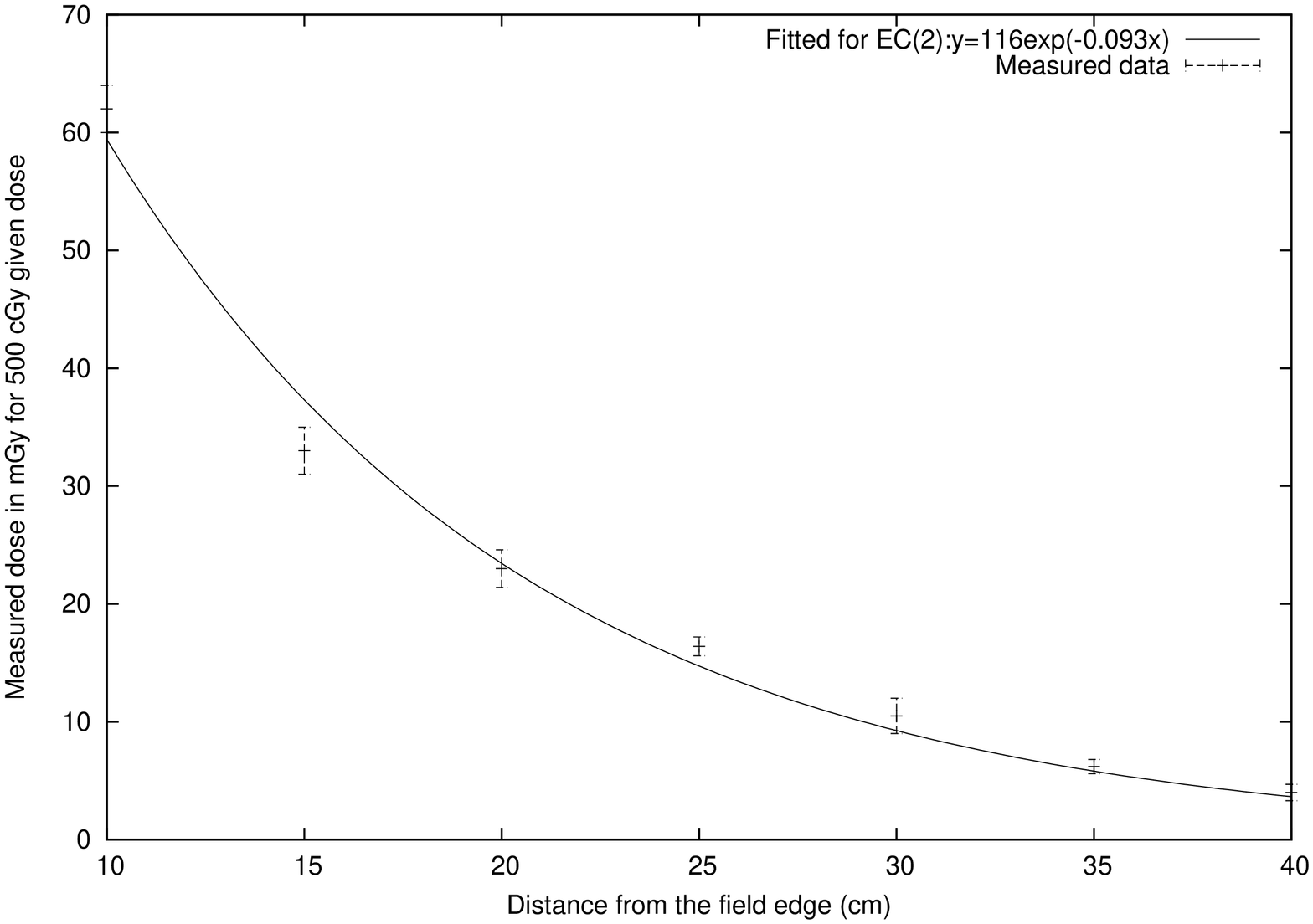}} \\
\subfloat[Field Size 14.14 cm $\times$ 7.07 cm]{\includegraphics[height=3 in, angle=270]{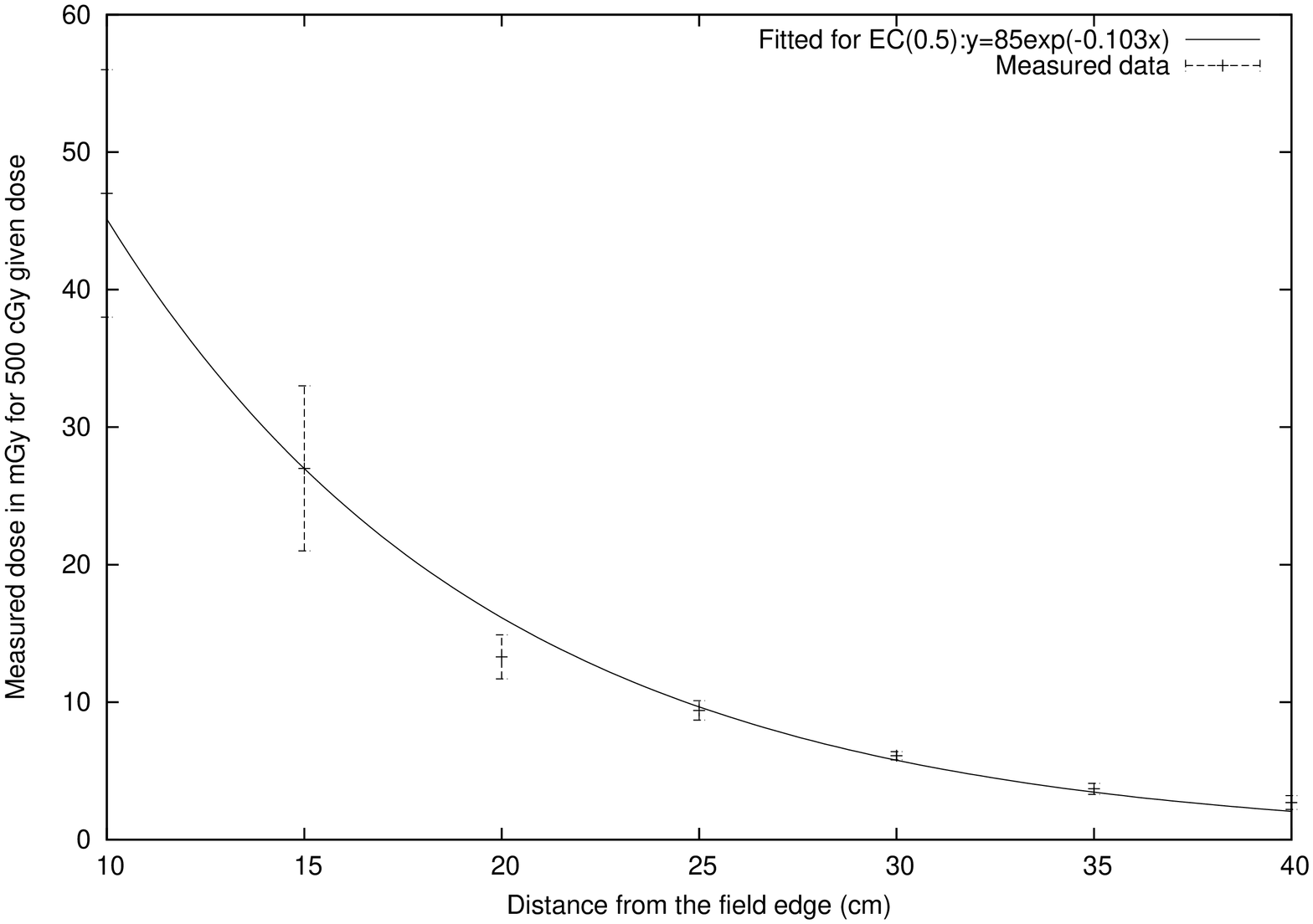}}
\subfloat[Field Size 20 cm $\times$ 5 cm]{\includegraphics[height=3 in, angle=270]{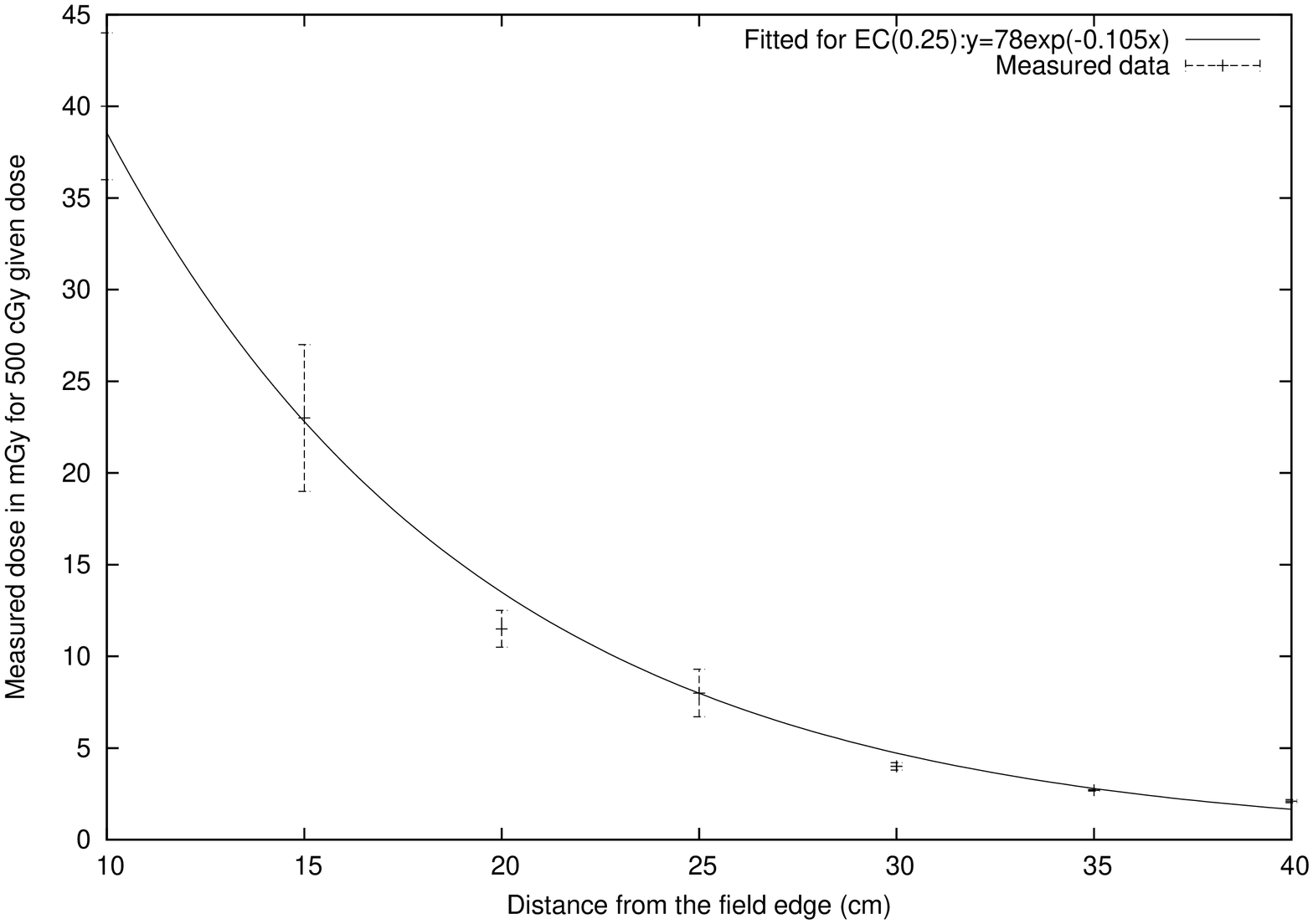}}
\caption{Dose distribution outside the beam with different Elongation Coefficients (EC) [(a) EC=4; (b) EC=2; (c) EC=0.5; (d) EC=0.25] of a 100 cm$^{2}$ field area (rectangular fields) for the $^{60}$Co unit with 80 cm SSD; 5 cm depth and 500 cGy given dose.}
\label{fig:EcUND}
\end{figure}
\begin{figure}[h]
\centering
\subfloat[EC=4]{\includegraphics[height=3 in, angle=270]{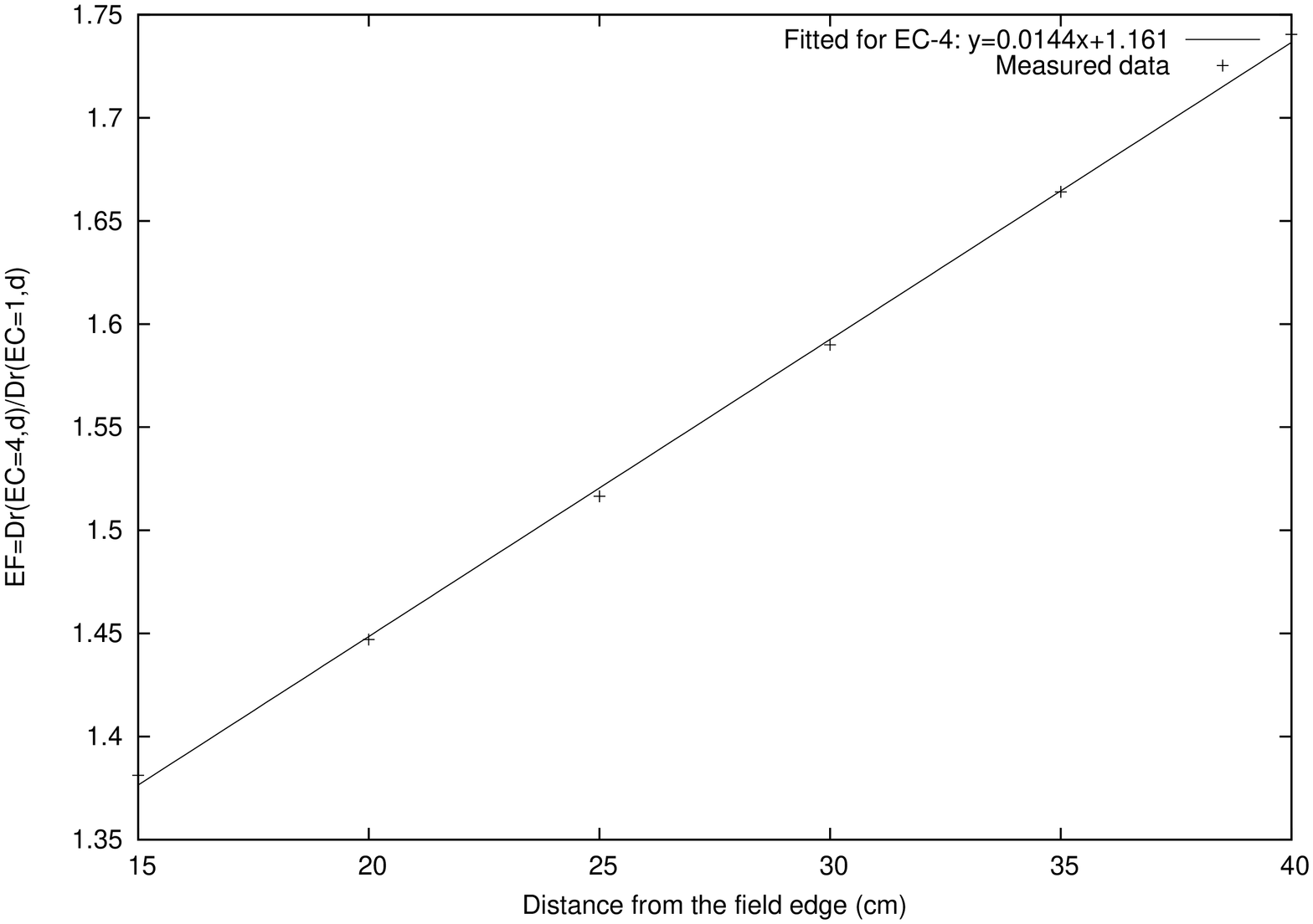}}
\subfloat[EC=2]{\includegraphics[height=3 in, angle=270]{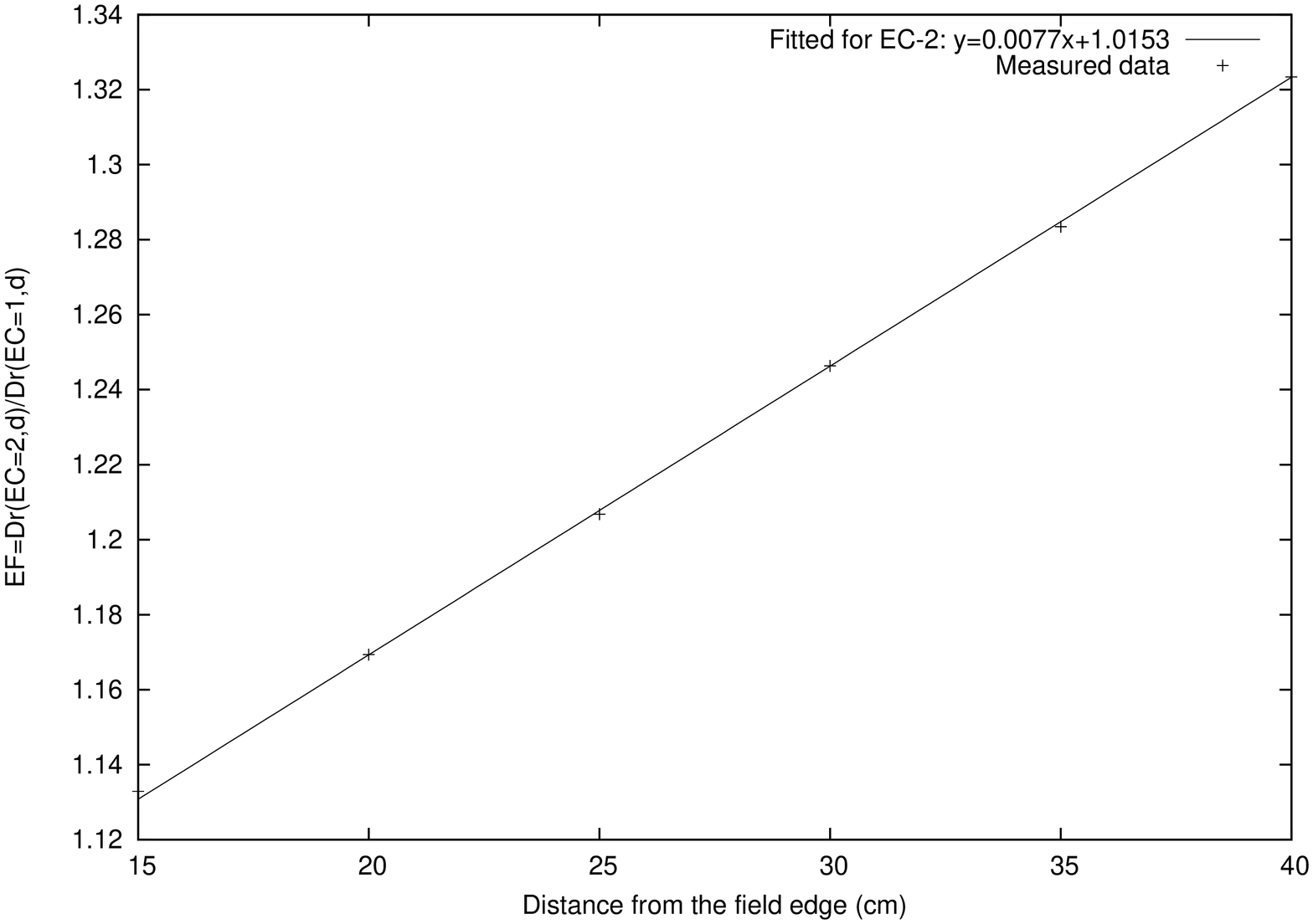}} \\
\subfloat[EC=0.5]{\includegraphics[height=3 in, angle=270]{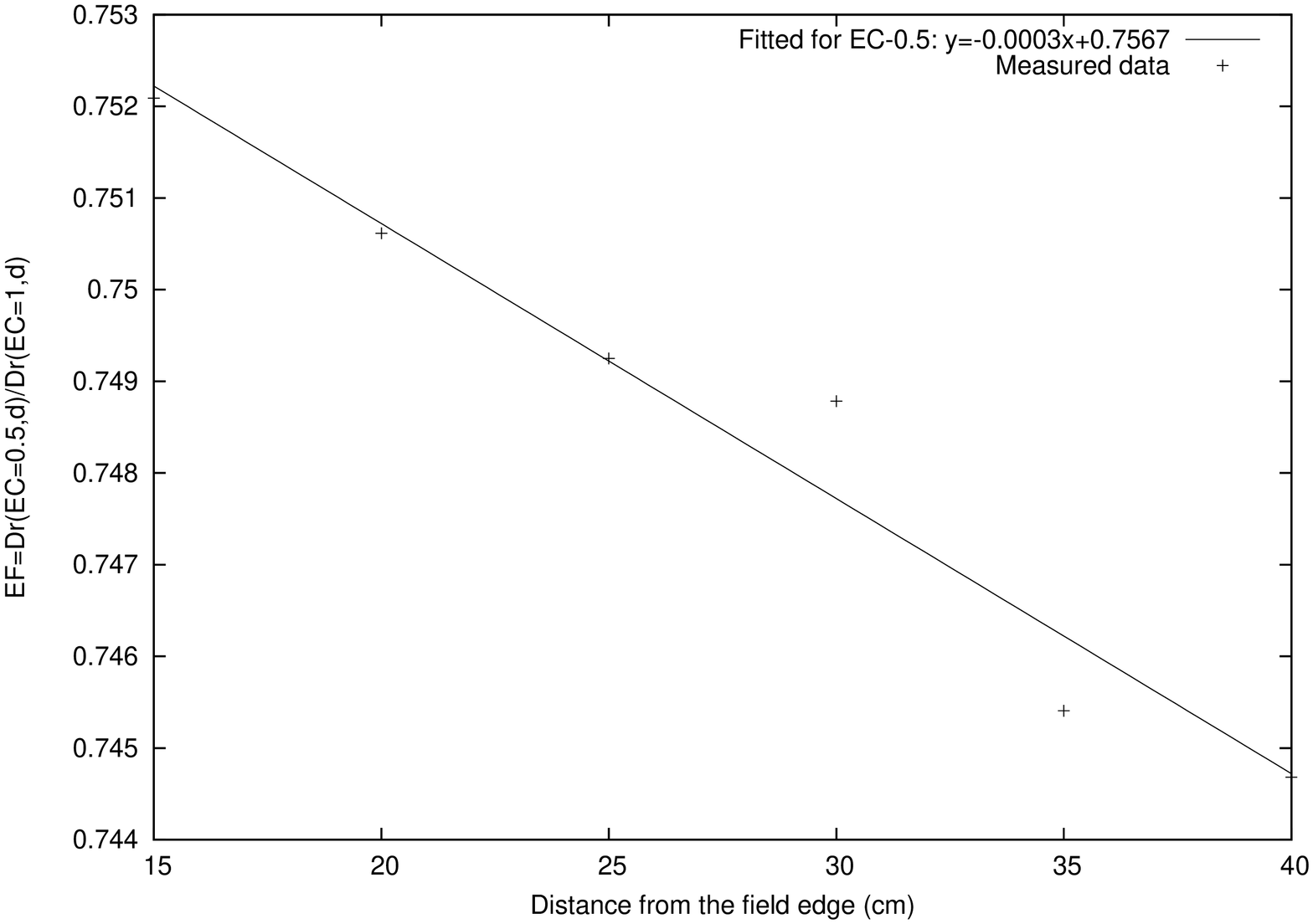}}
\subfloat[EC=0.25]{\includegraphics[height=3 in, angle=270]{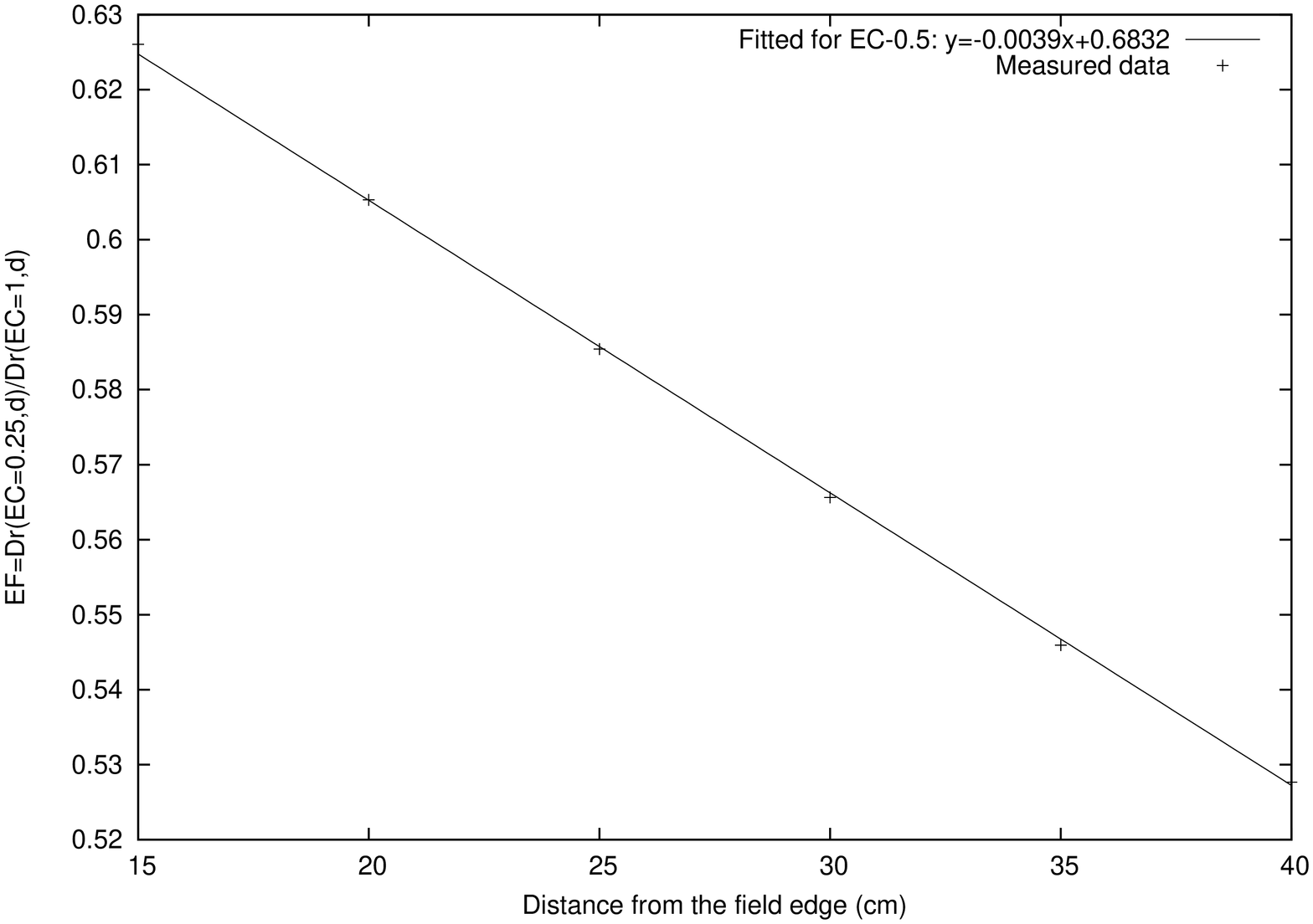}}
\caption{Variation of Elongation Factors (EF) with different Elongation Coefficients (EC) [(a) EC=4; (b) EC=2; (c) EC=0.5; (d) EC=0.25] of a 100 cm$^{2}$ field area (rectangular fields) for the $^{60}$Co unit with 80 cm SSD; 5 cm depth and 500 cGy given dose.}
\label{fig:EcUND1}
\end{figure}
Fig. 8 and Tale 2 represent the variation of radiation dose outside the treatment area at different depths with the field size (10 cm $\times$ 10 cm) for $^{60}$Co teletherapy. The data of Fig. 8 can be fitted closely to an exponential equation as $"y=a \times exp(-bx)"$. Using this fitted exponential equation, the data were recalculated to calculate the depth factors which are presented in Fig. 9. \\ \\
Each point of interest in the phantom contained 2 TLD chips except during the study of elongation factors, 3 chips were used to ensure a better accuracy for elongation factor. 
\begin{figure}[h]
\centering
\subfloat[Depth =1 cm]{\includegraphics[height=2 in, angle=270]{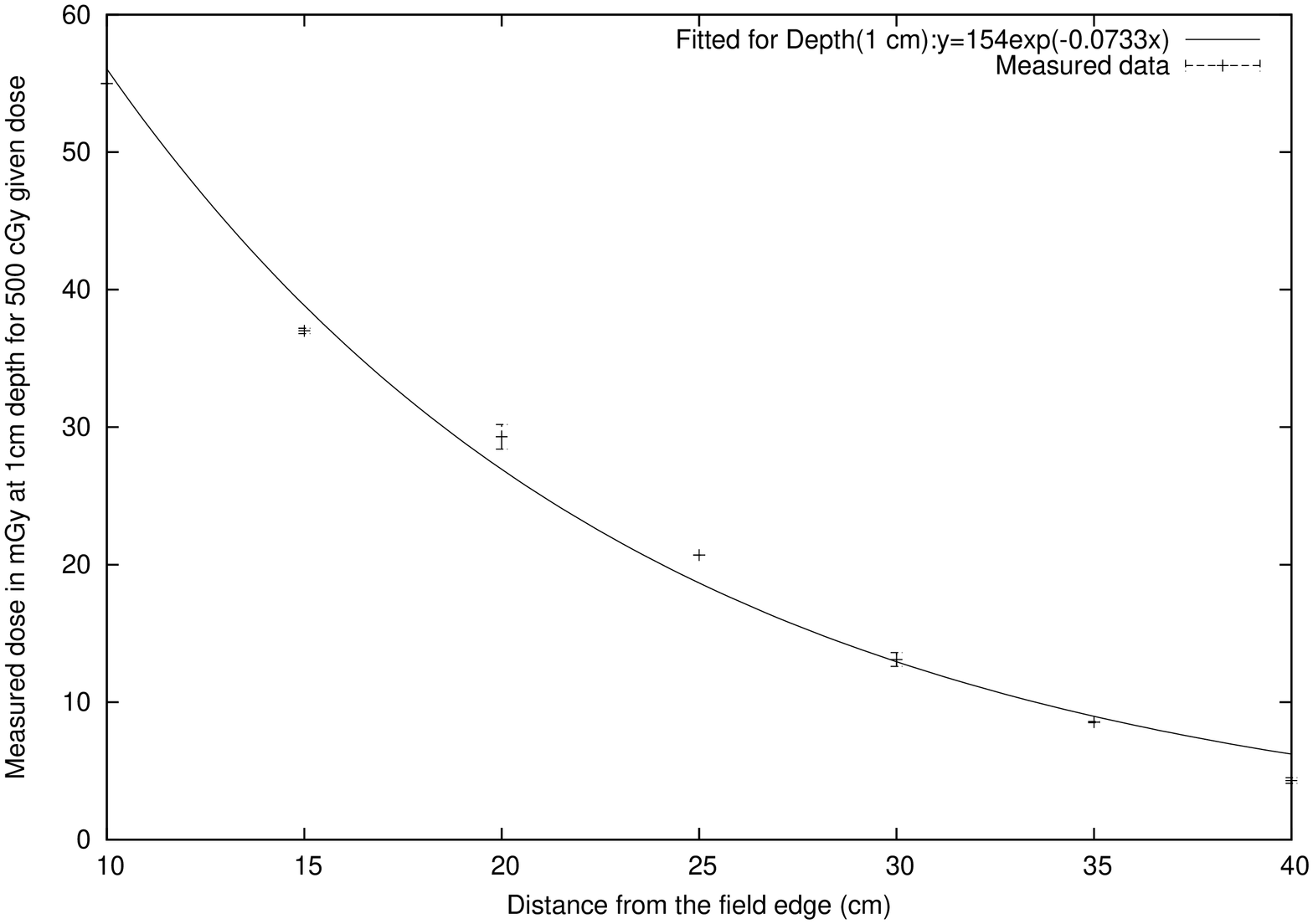}} 
\subfloat[Depth =5 cm]{\includegraphics[height=2 in, angle=270]{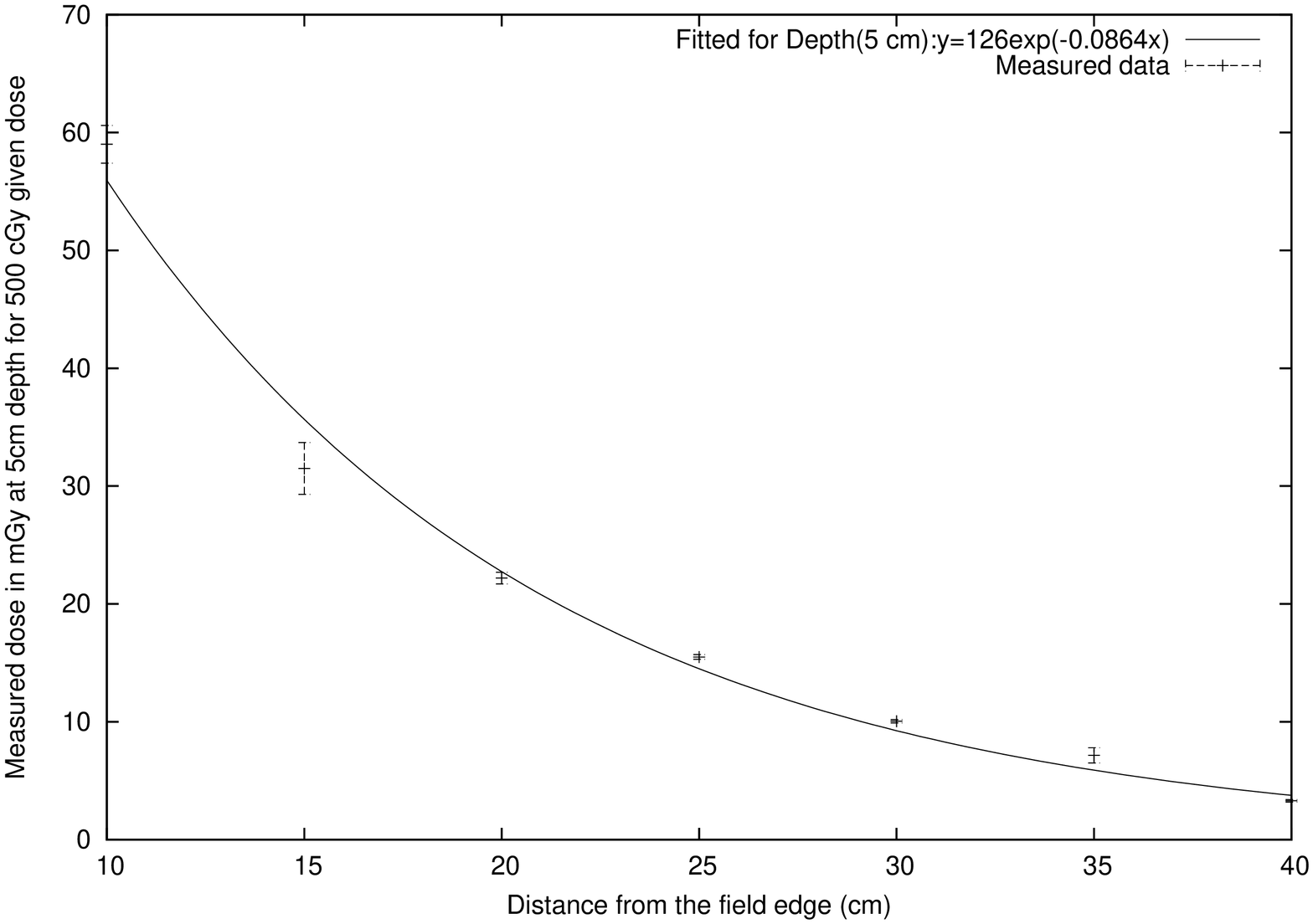}} 
\subfloat[Depth = 10 cm]{\includegraphics[height=2 in, angle=270]{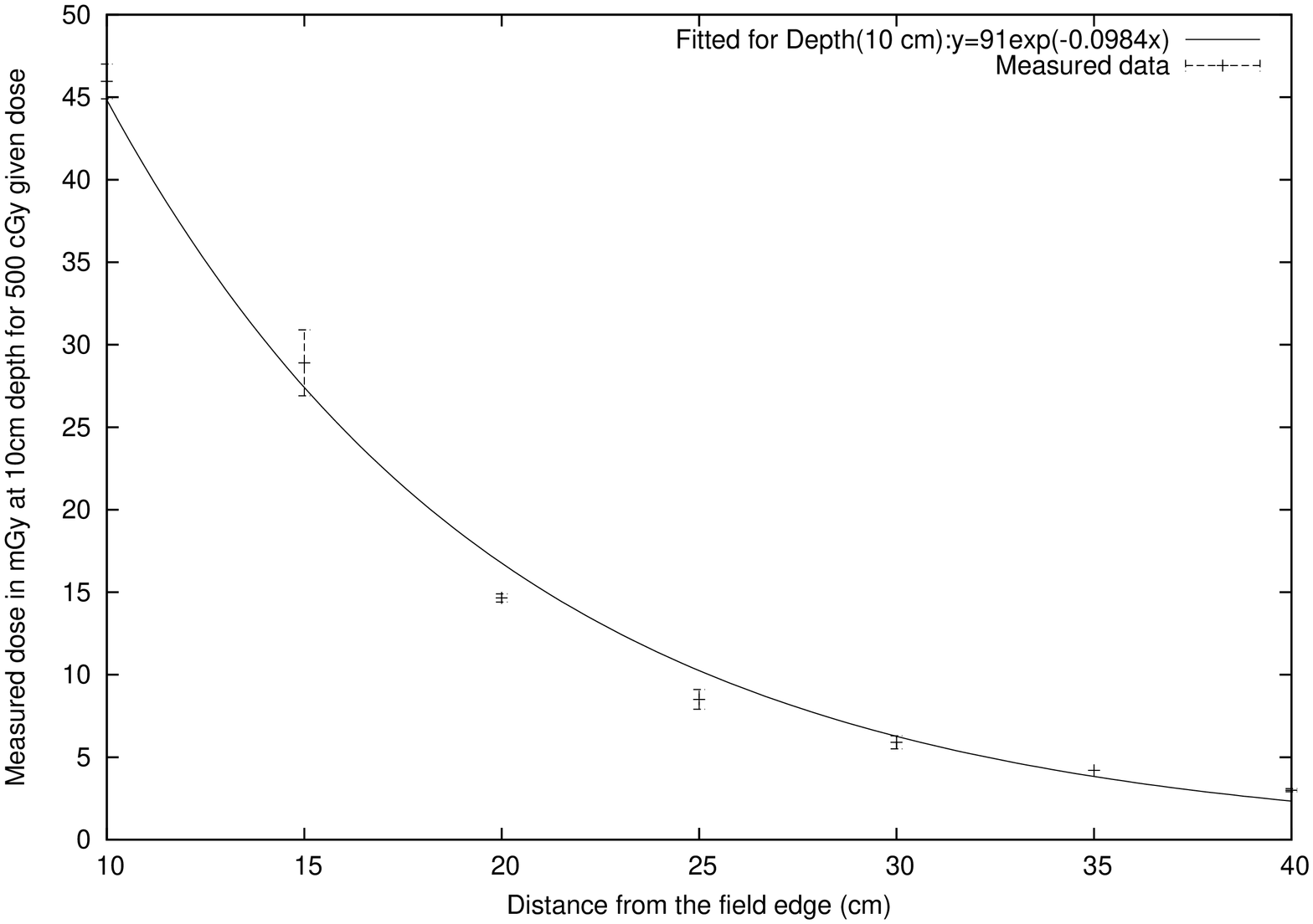}}
\caption{Dose distribution outside the beam with different depths for a 10 cm $\times$ 10 cm square field for the $^{60}$Co unit with 80 cm SSD and given dose 500 cGy.}
\label{fig:EcUND2}
\end{figure}
\begin{figure}[h]
\centering
\subfloat[$DF = \dfrac{Ds(z=1, d_{i})}{Ds(z=5, d_{i})}$]{\includegraphics[height=3 in, angle=270]{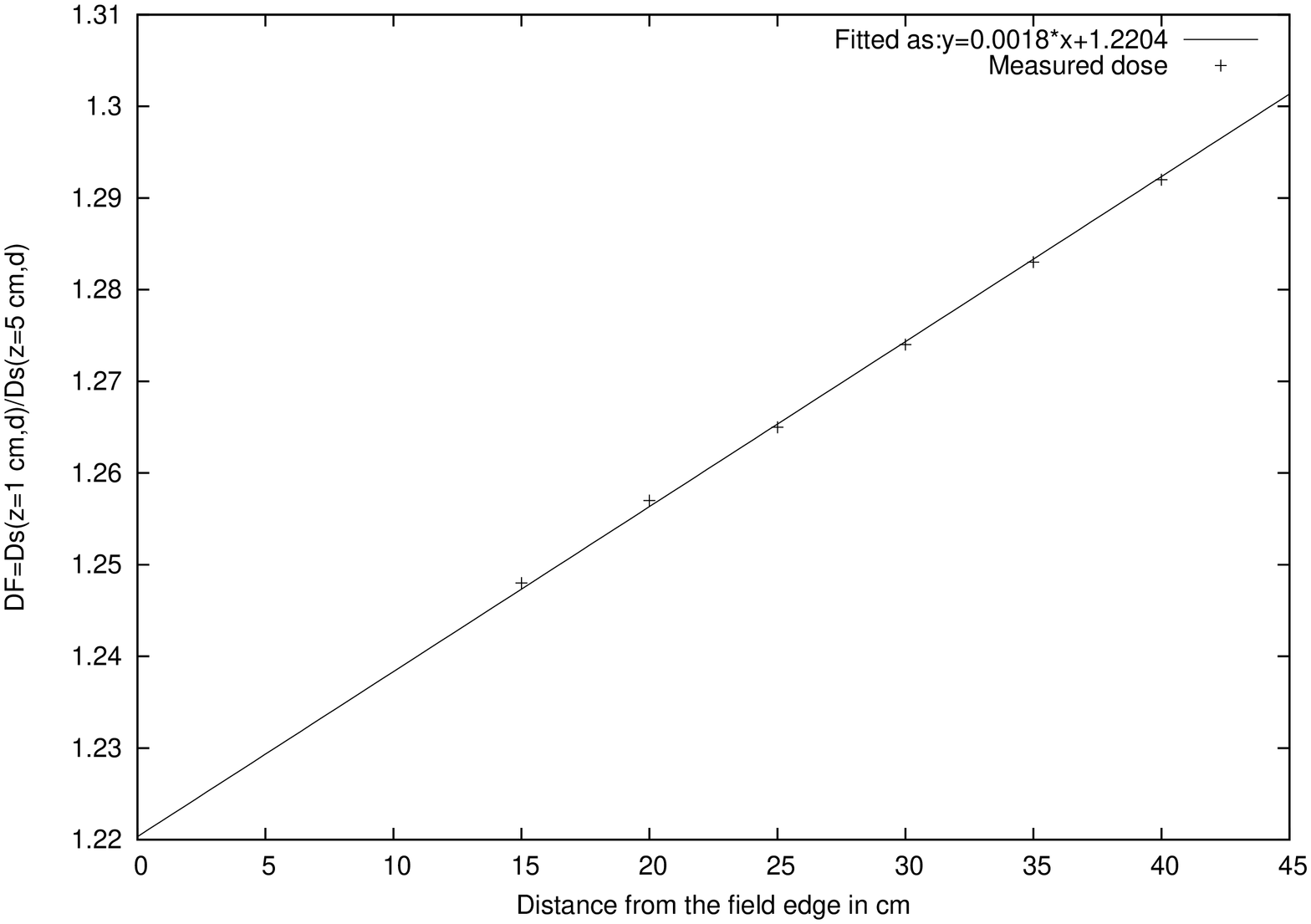}} 
\subfloat[$DF = \dfrac{Ds(z=10, d_{i})}{Ds(z=5, d_{i})}$]{\includegraphics[height=3 in, angle=270]{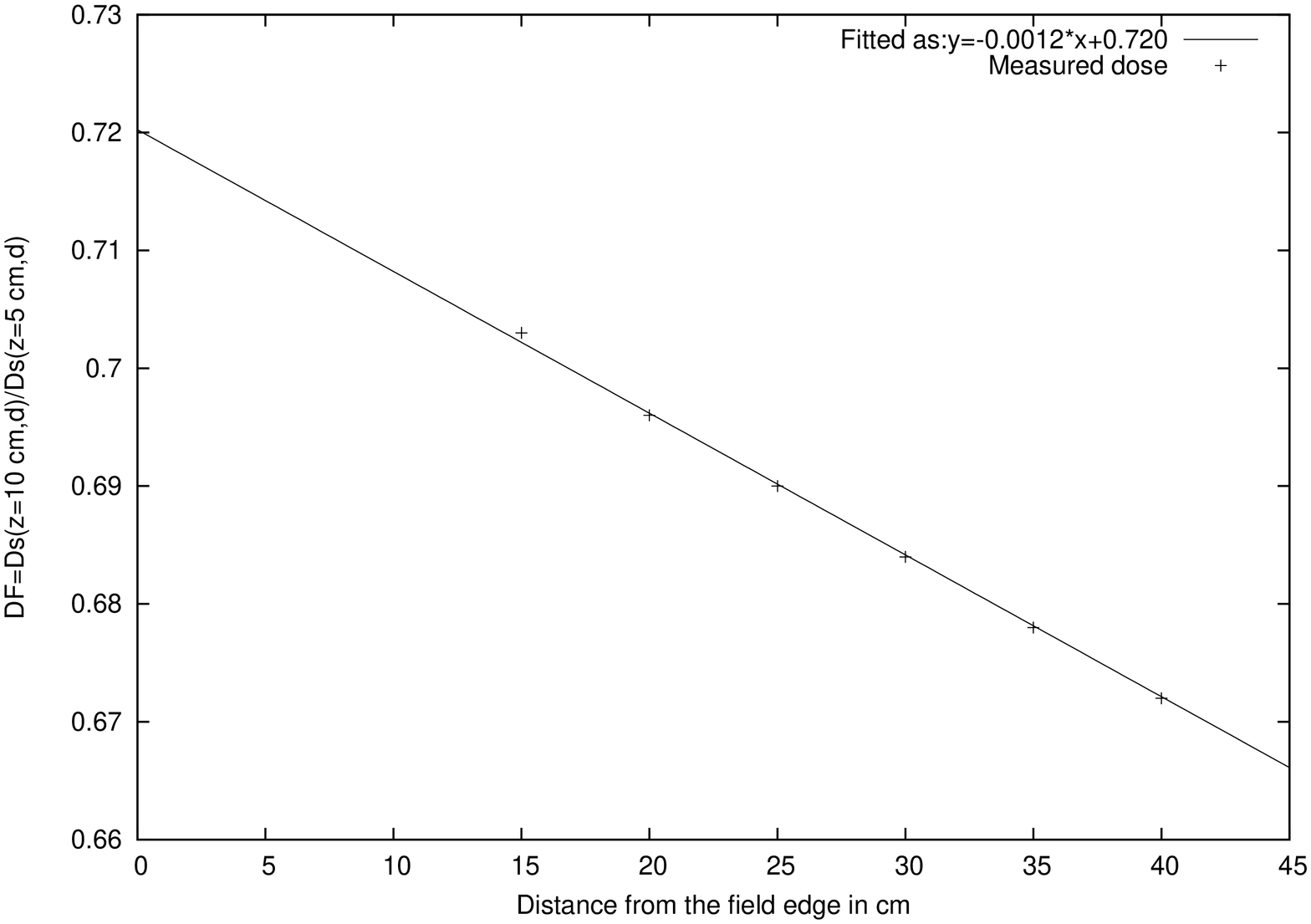}} 
\caption{Variation of Depth Factors (DF) with distance $(d_{i})$ for a 10 cm $\times$ 10 cm square field for the $^{60}$Co unit with 80 cm SSD.}
\label{fig:EcUND2}
\end{figure}
\section{DISCUSSIONS}
Radiation outside the treatment field arises from scatter generated by the patient, treatment couch, walls of 
the room and from leakage through the machine's shielding and collimators, which is known as secondary 
radiation. This secondary radiation is dependent on beam energy, field size, depth of measurement and 
distance from the field edge. Although there are differences in machine design which can affect the total secondary radiation, the measurements represented here (Figs. 3 - 5 ) are in good agreement with those reported by other authors (Ahmed et al., 1999; Fraass et al., 1983; Francois et al., 1988; Green et al., 1983; Green et al., 1985; Kase et al., 1983; Miah et al., 1998). \\ \\
As the field size increases, the fraction of dose contributed by phantom and wall scattered photons at a given distance from the field edge increases, eventually the fraction of dose collimator scatter and leakage decreases. So the total dose at any given point outside the treatment area increases with increasing field size, but the differences become less at large distances where collimator scatter and leakage dominate the dose. The variation of the dose with the different field shapes is illustrated in Figs. (3 - 6). The most obvious feature of these curves is the increase of the dose at a given distance from the field edge with the increase of the field elongation coefficient. This can be explained by the variation of the different components of the total secondary/scattered dose. Near the field, the scattered radiation from the patient is predominant. As the distance from the field edge increases, the radiation scattered by the collimator, the transmission through the collimator and the machine leakage become dominant (Kase et al., 1983). The scattered radiation from the collimator is presumably coming from the face of the beam-defining jaw seen from the point of measurement. For this reason, the scatter contribution from the jaw at this point will increase with the irradiation area of that jaw. In the case of rectangular field size  (5 cm $\times$ 20 cm), there is a ratio of four between the edges of the field and eventually the increase of areas of the two pairs of defining jaws. It could explain the differences between doses at equal distance from the field edge on the two axes (Francois et al., 1988). Close to the field edge, the isodose curve has an elliptical shape which follows approximately the shape of the field (Francois et al., 1988). If the distance increases further, the isodose line becomes elliptic again but with the two axes inverted (Francois et al., 1988). In fact, the isodose is not really elliptical in the area where the beam-defining jaws cross, because there is a double attenuation through the collimator jaws. The elongation factor (EF) at a given distances increases as the elongation coefficient (EC) increases. Figs. 6 and 7 represent the expected result, which are in good agreement with Francois et al., (1988). \\ \\
The dose decreases with the depth for distances from the field edge (Fig. 8). Due to scarcity of the facilities, the depth could not be made more than 10 cm for the determination of the depth factor. Close to the field edge, where scatter from the patient dominates, the total secondary dose increases with depth because the ratio of the scatter-to-primary increases with depth (Francois et al., 1988). Away from the field edge, where collimator scatter and leakage dominate, the total secondary dose decreases with depth because of attenuation in tissues beyond the depth of the maximum dose. Again as the depth increases, the depth factor at a given distance decreases (Fig. 9). Figs. 8 - 9 are representing the expected result, which are in good agreement with Francois et al. 1988.\\ \\
A few general comments may be made. Due to the manual method of measurements, there are some minimum differences of the readings. However, it may be observed in the Tables that, the measured dose in mGy for each distance of the phantom were quite consistent providing confidence in the present measurements. 
\section{CONCLUSIONS}
The 1980 BEIR Committee report (National Academy of Sciences, 1980) published the data for estimation of the excess risks of producing secondary tumors among long-term survivors. If a 30 year-old male received 1 Gy to one of his lungs, the estimated excess risk for developing lung cancer within his remaining expected lifetime of 40.5 yr would be approximately 1 percent (Kase et al., 1983). So, the measurement of scattered radiation outside the treatment area is important for the long lived survivors. \\ \\
Although of considerable uncertainty, our interest is to estimate the risk to patients from a dose outside the treatment field, and the possible benefits of diminishing that dose. The dose distribution measured over different distance from the end of the treatment area of the phantom during radiotherapy is the first reported measurement of this type in Bangladesh. The data reported in this paper are believed to be quite accurate. The data so obtained may be employed to compute doses in the underlying tissue of the whole body and the other internal radiosensitive organs, viz. lens of the eye, thyroid, oesophagus, stomach, cervix, kidney, gonad, bone marrow etc., and recommend measures for their protection. This is, in fact, the central objective of the radiological practice. Such protective measures may be taken and in turn, the cancer patients can have relatively better health as long as they survive which is also one of the major objectives of the technological advances in the present day medical treatment. \\ \\
An estimated incidence of new cancer cases in Bangladesh is about 200,000 per year and an annual death rate is about 150,000 (Huq et al., 1992). This situation definitely warrants improvement and expansion of the facilities, like treatment planning, study of scattered dose, etc. No reported data in this connection was found in Bangladesh so far. These experiments are generally conducted by an Anderson phantom in developed countries. But for developing countries, it is very difficult to procure such an expensive phantom. So, this experiment was conducted by using the locally fabricated phantom made of polystyrene sheets. \\ \\
These type of measurement could be extended to other radiotherapy centres globally where expensive Anderson phantoms are unavailable.   
\newpage 
\section*{Acknowledgements}
The authors are indebted to Professor A. D. Stauffer, Department of Physics and Astronomy, York University, Toronto, Canada and Professor T. Kron, Department of Radiation Oncology, Newcastle Mater Misericordiae Hospital Waratah, NSW 2298, Australia for valuable comments and discussions. The authors expresses their heartfelt gratitude to the authority of Bangladesh University of Engineering and Technology for giving necessary permission and providing financial support for this research work. The authors are grateful to the scientific and technical personnel of Health Physics Division, Atomic Energy Centre, Bangladesh Atomic Energy Commission, Dhaka for their sincere cooperation. The authors are also thankful to the doctors, physicists, technical staff and nurses of the Delta Medical Centre Limited, Dhaka and Radiotherapy Department, Dhaka Medical College and Hospital, Dhaka, Bangladesh for their friendly and sincere cooperation. Finally, one of the authors (M.F. Ahmed) is thankful to York University, Toronto, Canada for providing help during the writing this manuscript.     
\section*{References} 
\indent  Ahmed, M. F., Begum, Z., Miah, F. K., and Chowdhury, Q., Dose distribution over different organs of some retinoblastoma cancer patients undergoing radiotherapy, Journal of Medical Physics, 24 (4), 190 - 194 (1999). \\ \\
\indent Ahmed, M.F., Distribution of Exposure Doses Over Various Parts of the Body of Cancer Patients Due to Different Radiotherapy Procedures in Bangladesh, M.Sc. Thesis, University of Dhaka (1994).\\
\\
\indent Ahmed, M. F., Dose distribution to organs outside (Laterally and Depthwise) the treatment area following elongation co-efficient in case of radiotherapy treatment, M. Phil. Thesis, Bangladesh University of Engineering and Technology (2000). \\
\\
\indent Brandan, M. E., Perz Pastenes, M. A., Ostrosky-Wegman, P., Gonsebatt, M. E., Diaz-Perches, R., Mean dose to Lymphocytes During Radiotherapy Treatments, Health Phys., 67, 326-329 (1994). \\ 
\\
\indent Fiaz, M. Khan, \textit{The Physics of Radiation Therapy,} Second Edition, Williams and Wilkins, USA, pp. 176 - 177 (1994). \\ \\
\indent Fraass, B., and Van De Geijn, J., Peripheral dose from megavolt beams, Med. Phys., 10, 809 (1983). \\ 
\\
\indent Francois, P., Beurtheret, C. and Dutreix, A., Calculation of the dose delivered to organs outside the radiation beams, Med, Phys. 15(6), 879 (1988) \\ 
\\
\indent Goil De Planque Burke, IEEE, Thermoluminescence Dosimetry, Trans. Nucl. Sci., NS-23, 1224(1976). \\
\\
\indent Gottfried, B. S. (1979) \textit{Theory and problmes of Introduction to Engineering Calculation}, pp. 55-59, Schaum's outline series, McGrawHill Book Co., New York.\\ \\
\indent Green, E. D., Chu, G. L. and Thomas, D., Dose level outside radiotherapy beams, Br. J. Radiol., 56, 543 (1983). \\
\\
\indent Green, E. D., Karup, P., Sims, C. and Taylor, R., Dose levels outside radiotherapy beams, Br. J. Radiol., 58 453 (1985). \\
\\
\indent Hussain, S. R., Molla, M. A. R., Rahman, M. M., Alam, M. N., Hai, M. A. and Abdullah, S. A., Assessment of occupational exposure and provision of recommendations and guidance on measures for the radiological protection of occupationally exposure persons in Bangladesh, Final Report, IAEA/BAEC Contact No. 2000/RB, Atomic Energy Centre, Dhaka, Bangladesh (1982). \\
\\
\indent ICRP (International Commission on Radiological Protection). ICRP publication 44, Pergamon Press, Oxford (1985). \\ \\
\indent Kase, K., Svensson, G. Wolbarst, A., and Marks, M., Measurement of dose from secondary radiation outside a treatment field, Int. J. Radiat. Oncol. Biol., 9(8), 1177-1183 (1983). \\
\\
\indent Miah, F. K., Ahmed, M. F., Begum, Z., Alam, B. and Chowdhury, Q., Dose distribution over different parts of cancer patients during radiotherapy treatment in Bangladesh, Radiation Protection Dosimetry, 77(3), 199-203 (1998). \\
\\
\indent National Academy of Sciences, Committee on the Biological Effects of Ionizing Radiations (BEIR): The effects on populations of exposure to low levels of ionizing radiation: 1980. In: K. Kase, G. Svensson, A. Wolbarst, and M. Marks, "Measurement of dose from secondary radiation outside a treatment field, "Int. J. Radiat. Oncol. Biol. 9(8), 1177 - 1183 (1983).\\ 
\\
\indent Schulz, R. J. and Nath, R., Med. Phys. 6, 153 (1979). \\
\\
\indent Sherazi, S. and Kase, K., Measurements of dose from secondary radiation outside a treatment field; Effects of Wedges and blocks, Int. J. Radiat. Oncol. Biol. Phys., 11, 2171 (1985). \\
\\
\indent $^{90}$Sr/$^{90}$Y-Irradiator Calibration Certificate, Certified by Dr. Jurgen Fellinger, Radiation Protection Physicist, Bicron Technologies (1995). \\ \\
\indent Starkeschall, G., George, F. St. and Zellmer, D., Surface dose for megavoltage photon beams outside the treatment field, Med. Phys., 10, 906 (1983). \\ \\
\indent User's Manual, Model 3500 Manual TLD Reader, Harshaw Bicron Radiation Measurement Products, Publication No. 3500-0-U-0793-005, (1993).\\ \\
\end{document}